\documentclass[11pt,a4paper]{article}
\usepackage{jheppub} 
\usepackage[T1]{fontenc}
\usepackage[utf8]{inputenc}
\DeclareUnicodeCharacter{2019}{\ifmmode^{\prime}\else\textquoteright\fi}

\usepackage{caption}
\pdfoutput=1 

\usepackage{xcolor,bm,tikz,pgfplots}
\usepackage{enumerate}
\usepackage{framed}
\usepackage{mdframed}
\usepackage{amsmath}
\usepackage{amsfonts}
\usepackage{mathrsfs}
\usepackage{amssymb}
\usepackage[normalem]{ulem}
\usepackage{graphicx, rotating}
\usepackage{epsfig}
\usepackage{latexsym}
\usepackage{graphicx}
\usepackage{color}

\usepackage{slashed}
\usepackage{hyperref}
\usepackage{epstopdf}
\usepackage[title]{appendix}
\usepackage[font=footnotesize,labelfont=]{caption}
\AppendGraphicsExtensions{.tif}
\hypersetup{colorlinks=true, citecolor=bluscuro, linkcolor=black, urlcolor=bluscuro}
\definecolor{rossos}{cmyk}{0,1,1,0.55}
\definecolor{bluscuro}{rgb}{0.15, 0.2, .85}
\definecolor{bluchiaro}{cmyk}{1,.3,0.,0.1}
\definecolor{Green}{rgb}{0, 0.65, 0.31}

\newcommand{\be}{\begin{equation}}
\newcommand{\ee}{\end{equation}}

\newcommand{\rd}{{\rm d}}
\newcommand{\ii}{\mathrm{i}}

\newcommand{\vev}[1]{\bigl\langle#1\bigr\rangle}

\title{\boldmath Tracing out massive fields in cosmology}

\author[a]{Guanhao Sun}
\author[b]{and Sam S. C. Wong}

\affiliation[a]{Department of Physics and Chongqing Key Laboratory for Strongly Coupled Physics, Chongqing University, Chongqing 401331, People’s Republic of China}

\affiliation[b]{Department of Physics, City University of Hong Kong, Tat Chee Avenue, Kowloon, Hong Kong SAR, China}

\emailAdd{sunguanhao@cqu.edu.cn}
\emailAdd{samwong@cityu.edu.hk}

\abstract{
Cosmological correlators are calculated through the in-in formalism in (quasi-)de Sitter background. Its effective description is best represented by a reduced density matrix. In this work, we study how integrating out a massive field in de Sitter space determines the reduced density matrix of a light field at a finite observation time. Starting from a quasi-single-field model with non-linear couplings, we first evaluate the massive field wavefunctional at fixed final value and then trace over this value. The trace generates branch-mixing terms which are generally present in the Schwinger--Keldysh description. For a conformally coupled massive field, we show explicitly that the light field bispectrum from the effective description agrees with a full theory calculation. This example also shows that decay of a field at the future boundary does not by itself justify discarding its contribution to the trace. We discuss the differences between an exact nonlocal effective description and the local effective field theory obtained through a large mass expansion. We then identify which local cubic terms affect the late-time probability and which contribute only a wavefunctional phase. Finally, we organize the resulting kernels by response, noise, and Schwinger--Keldysh consistency conditions, and discuss the symmetry constraints on the reduced density matrix. 
}


\begin{document}
\maketitle
\flushbottom 
\clearpage

\section{Introduction}   \label{sec:intro}
Cosmological correlation functions can preserve information about massive fields that decay at late times \cite{Chen2009we,Chen2009zp,ArkaniHamed2015bza,Chen2016QuantumClocks}. Their behaviors are important for observational works such as galaxy-bias modeling and forecasting~\cite{Sefusatti2012QSF,Meerburg2017Prospects,Dizgah2018GalaxyBispectrum,Dizgah2018ScaleBias,Assassi2015Bias,Dizgah2018LongLived}, primordial non-Gaussianity constraints \cite{Planck2018NG}, and cosmological collider searches \cite{Sohn2024Collider,Suman2025ColliderII,Cabass2025BOSS,Kumar2026ScalarSearch,Philcox2026ScalarCMB}. When these heavy fields are unobserved, their effects are encoded in the reduced density matrix of the light sector. How should we construct this reduced description to reproduce finite-time in-in correlators, and when is the final-time trace essential even when the massive field decays toward the future boundary? We address both questions in a two scalar model.

Reduced dynamics can be expressed through an influence action on the two Schwinger--Keldysh histories~\cite{Schwinger1961Brownian,FeynmanVernon1963,Keldysh1965,Jordan1986Causal,CalzettaHu1987CTP,Weinberg2005vy}. Recently its applications in cosmology have been studied in a series of works on the effective description of inflation as an open system, including dissipation and noise, quantum decoherence, and resummation of the late time evolution~\cite{LopezNacir2012Dissipative,Burgess2015Open,Boyanovsky2015Effective,Boyanovsky2016Stochastic,Burgess:2022nwu,Colas2022Open,AguiSalcedo2024Open,Burgess:2024eng,ColasQinTong2026Collider,Colas2025Lectures,Pajer2026Lectures}. The top-down procedure where one traces out a common final field value has also been studied~\cite{GreenSun2025,Cespedes2025Quartic,CespedesColas2026}. Integrating or tracing out the heavy degrees of freedom does not require $m\gg H$, but a local effective field theory (EFT) description requires a physical scale hierarchy and adiabatic evolution \cite{Cespedes2012hu,Achucarro2012sm,Achucarro2012Decoupling,DuasoPueyo2026Asymptotic}. When $m\sim H$, the effective description generally remains nonlocal.

Calculations of cosmological correlators make use of a variety of theoretical tools, including diagrammatic expansions, the wavefunctional method, bootstrap, and cutting rules~\cite{Chen2017ryl,ArkaniHamed2018kmz,Goodhew2020hob,Baumann2022jpr,GoodhewJazayeriLeePajer2021,Cespedes2020Time,Aoki:2023wdc,AguiSalcedo2023Analytic,Werth2023Flow,Qin2023Seeds,Xianyu:2023ytd,Aoki:2024uyi,Qin:2025xct,Palma2026Map,Melville:2021lst,Jazayeri:2021fvk,Stefanyszyn:2024msm}. In this work, we start with the wavefunctional of a system with a light field derivatively coupled to a self-interacting massive field, then derive the reduced density matrix for the light field at finite time. Our model falls into the category of quasi-single field inflation. See~\cite{Chen2012LargeMass,Assassi:2012zq,NoumiYamaguchiYokoyama2013,GongPiSasaki2013Equilateral,Iyer2018PartialEFT,TongWangZhou2018Warm,Huenupi:2026abj,Huenupi:2026aqc,Wang:2026lff,Pinol:2026xnl} for some related studies.  

Working in the Schwinger-Keldysh setup to calculate correlators is different from calculating scattering amplitudes in the usual in-out setup. For example, total derivative terms and terms proportional to the equation of motion become important for in-in calculations~\cite{Braglia2024Derivatives,Kawaguchi:2024lsw}. The EFT for correlators is therefore different from the EFT for scattering amplitudes~\cite{GreenSun2025}. Simply doubling the EFT for scattering amplitudes as an in-in EFT would certainly generate discrepancies between the EFT and the full theory~\cite{An:2017hlx,Tong:2017iat}. Specifically, the homogeneous part of the solution to the equation of motion of the heavy field encodes its boundary value. This part is usually suppressed in the EFT for scattering amplitudes where we have both asymptotic in and out states. Tracing over (integrating out) the heavy field then identifies its boundary value at final time on the two branches. In doing so, we generate cross branch mixing in the effective description, which is beyond single branch scattering amplitude EFT. We formalize this procedure in this work.

After working out the general framework, we explicitly check the conformally coupled case, $m^2=2H^2$, against a direct in-in calculation in the original two-field theory. Although the massive field decays toward the future boundary, we see corrections introduced by the trace. We also show that for a generic mass, our prescription recovers the squeezed limit behavior of the full theory. Lastly, we check that in the heavy mass limit we obtain a local EFT expansion. We then match the response and noise kernels and discuss how symmetries constrain the reduced density matrix.

This article is organized as follows. Section~\ref{sec:general} gives the general finite-time construction. Section~\ref{sec:qsfi} computes the bispectrum kernels and their conformal, squeezed, and heavy-mass limits. Section~\ref{sec:bottom-up-open-eft} organizes the response, noise, and symmetry constraints, and Section~\ref{sec:conclusion} discusses the implications. Appendix~\ref{app:exact-three-point} collects the mode functions and explicit checks.

Throughout the calculation, $\zeta$ denotes a canonically normalized light scalar. It can be promoted to the curvature perturbation provided that we carefully carry out inflationary normalization and nonlinear symmetry completion discussed in various works~\cite{Cheung2007st,Maldacena2002vr,Creminelli2012ed,Pimentel2013gza,Green:2020ebl}. Our model uses $a^2g\sigma\zeta'$ mixing with constant $g$, corresponding to a cosmic-time mixing coefficient $g/a$. Its time dependence affects both the momentum scaling and the local phase cancellations. In Appendix~\ref{app:a3-mixing}, we compare it with $a^3g\sigma\zeta'$ mixing, whose cosmic-time coefficient is constant. 

\section{General procedures: integrating out heavy fields and the reduced density matrix at finite time} \label{sec:general}
In this section, we outline the general procedure for constructing a top-down effective description of in-in correlators by integrating out a massive field. The primary quantity of interest is the reduced density matrix. On a Schwinger--Keldysh contour, tracing out degrees of freedom generally introduces non-factorizable cross-terms coupling the forward ($+$) and backward ($-$) contours ~\cite{FeynmanVernon1963,GreenSun2025}. Unlike the EFT for scattering amplitudes, these cannot be represented by a product of light-field wavefunctionals, $\Psi^*[\zeta_-]\Psi[\zeta_+]$, so a single-branch action alone does not specify the full reduced theory. 

Consider the following interacting theory with a massless field $\zeta$ and a massive field $\sigma$ in de-Sitter spacetime,
\begin{align}
    S[\zeta,\sigma ] = \int d^4x \sqrt{|g|} \left[  -\frac{1}{2} \partial_{\mu}\zeta \partial^{\mu}\zeta  -\frac{1}{2} \partial_{\mu}\sigma \partial^{\mu}\sigma  - \frac{1}{2}m^2 \sigma^2 
    +  {\cal L_{\rm int}}[\sigma, \sigma', \zeta, \zeta']  \right] \ .
\end{align}
Here a prime denotes a derivative with respect to conformal time $\tau$, and ${\cal L}_{\rm int}$ includes self-interactions and mixed interactions between the two fields. We will construct the tree-level effective description of $\zeta$ by explicitly integrating out $\sigma$, and identify the cross-branch terms required by the in-in formalism \cite{Weinberg2005vy,Chen2017ryl}.

Start with the in-in path integral calculation of the expectation value of some operator $\hat{\cal{O}}(\tau_0)$,
\begin{align}
    \vev{\hat{\cal O}(\tau_0)} = \int\!{\cal D}\bar\Phi\int_{\Phi_+(\tau_0)=\Phi_-(\tau_0)=\bar{\Phi}} \mathcal{D}\Phi_+ \mathcal{D}\Phi_- \, \mathcal{O}[\bar\Phi] \, e^{iS[\Phi_+]-iS[\Phi_-]}\rho_i[\Phi_+(\tau_i),\Phi_-(\tau_i)] \ .
\end{align}
Here we use $\Phi$ to represent a general field multiplet. If the initial state is the vacuum $|\Omega \rangle$, it further simplifies to 
\begin{align}
     \vev{\hat{\cal O}}  = \int \mathcal{D}\bar{\Phi} \, \left|\Psi\left[\bar{\Phi}\right] \right|^2 \mathcal{O}\left[\bar{\Phi}\right] \ ,
\end{align}
where 
\begin{align}
    \Psi\left[\bar{\Phi}\right]  ={\cal N} \int_{\Phi(\tau \to -\infty(1-i\epsilon)) = 0}^{\Phi(\tau_0) = \bar{\Phi}} \mathcal{D}\Phi \, \exp\left( iS[\Phi] \right)
\end{align}
is the vacuum wavefunctional at time $\tau_0$. The early-time contour $\tau\to-\infty(1-i\epsilon)$ selects the positive frequency Bunch--Davies solution. We assume this initial state for the rest of the work.

At tree level, one calculates the wavefunctional by evaluating the action on the solution to the classical equations of motion. Take the equation
\begin{equation}
    (m^2 - \Box) \Phi = J[\Phi] \ .
\end{equation}
The solution should satisfy the boundary conditions $\Phi(\tau_0) = \bar\Phi$ and $\Phi(\tau \rightarrow -\infty) \sim \exp(iE\tau)$, formally written as 
\begin{equation}
    \Phi_k(\tau) = \bar\Phi_k K_k(\tau) + \int d\tau'\,a^4(\tau')G_k(\tau,\tau')J_k[\Phi](\tau') \ ,
\end{equation}
where $K_k$ satisfies $(m^2-\Box)K=0$ and $G_k$ obeys a vanishing Dirichlet boundary condition at $\tau_0$. We use $G$ as the inverse of $m^2-\Box$. Suppose the mode function of the field is $u_k(\tau)$, one can write 
\begin{align}
    K_k(\tau)&=\frac{u_k(\tau)}{u_k(\tau_0)},
    \quad
    P_k(\tau_0)=|u_k(\tau_0)|^2\ ,  \nonumber \\ 
    G_k(\tau,\tau') &=  iP_k\left[\theta(\tau-\tau')K_k(\tau')K_k^*(\tau) + \theta(\tau'-\tau)K_k(\tau)K_k^*(\tau') -K_k(\tau)K_k(\tau') \right] \ . 
\end{align}
Plugging these back into the action, one can calculate the wavefunctional of the theory perturbatively. Lastly, we interpret the modulus square of the wavefunctional as the probability distribution, and calculate the expectation values perturbatively. 

On the other hand, the path integral on the two branches can be interpreted as the time evolution from the initial to the final state. We can then write the final density matrix as
\begin{equation}
    \rho[\bar\Phi(\tau_0),\tilde\Phi(\tau_0)] = \int_{\Phi_+(\tau_0) = \bar\Phi, \Phi_-(\tau_0) = \tilde\Phi} {\cal{D}}\Phi_+ {\cal D} \Phi_- \exp(iS[\Phi_+]-iS[\Phi_-]) \ .
\end{equation}
For a pure state, $\rho[\bar\Phi,\tilde\Phi]=\Psi[\bar\Phi]\Psi^*[\tilde\Phi]$, so that its diagonal is $|\Psi[\bar\Phi]|^2$, and $\langle\hat{\cal O}(\tau_0)\rangle=\operatorname{Tr}[\rho(\tau_0)\hat{\cal O}(\tau_0)]$. The path-integral, wavefunctional, and density-matrix descriptions of in-in correlators are equivalent \cite{Palma2026Map}.

In our two-field example, $\Phi=(\zeta,\sigma)$. In spirit of effective field theory, if the measured operators involve only $\zeta$, we can marginalize over $\sigma$ and construct an effective description of $\zeta$ only. This process must be carried out on the full Schwinger--Keldysh contour and involves an integral over the boundary value $\bar\sigma = \bar\sigma_+ = \bar\sigma_-$ at the future boundary \cite{GreenSun2025,Cespedes2025Quartic}. Simply doubling an in-out bulk EFT action generally misses part of this matching. A mixed reduced state cannot be represented by a single effective light-field wavefunctional, and its reduced density matrix is the appropriate object. Therefore, strictly speaking there is no ``effective wavefunctional''. With these in mind, we proceed to the steps for the top-down construction of the $\zeta$ effective theory using a reduced effective density matrix.

Write the massive field equation as $(m^2 - \Box)\sigma=J$, with $J(x)=[\sqrt{-g(x)}]^{-1}\delta S_{\rm int}/\delta\sigma(x)$. At fixed $\bar\sigma$, the two solutions have the form
\begin{align}
    \sigma_{k,+}(\tau)
    &=\bar\sigma_k K_k(\tau)
      +\int_{{\cal C}_+}d\tau'\,a^4(\tau')G_k(\tau,\tau')J_{+,k}(\tau')\ ,
    \label{eqn:simga_+_formal_sol}\\
    \sigma_{k,-}(\tau)
    &=\bar\sigma_k K_k^*(\tau)
      +\int_{{\cal C}_-}d\tau'\,a^4(\tau')G_k^*(\tau,\tau')J_{-,k}(\tau')\ .
    \label{eqn:simga_-_formal_sol}
\end{align}
The differences come from different choices of the $i\epsilon$ sign in the two branches. The normalized homogeneous solution obeys $K_k(\tau_0)=1$, while $G_k$ is a Green's function with a Dirichlet condition at $\tau_0$. 

Next, we substitute the classical solutions of equations~\eqref{eqn:simga_+_formal_sol}  and \eqref{eqn:simga_-_formal_sol} at fixed $\bar\sigma$ into the action, then integrate over the  boundary conditions $\bar\sigma$. At tree level, stationarity of the boundary integral $\int{\cal D}\bar\sigma\,e^{i(S_+-S_-)}$ imposes
\begin{align}
    \frac{\delta(S_+^{\rm os}-S_-^{\rm os})}{\delta\bar\sigma} = 0 \ .
    \label{eq:general-momentum-sewing}
\end{align}
For an action with at most first derivatives, this amounts to $\bar \pi_{\sigma,+}-\bar \pi_{\sigma,-}=0$ at tree level~\cite{Galley2013Nonconservative,GalleyTsangStein2014}, where $\bar\pi_\sigma$ is the canonical momentum at the boundary $\tau_0$. The trace therefore identifies both the heavy-field values and their momenta at the final surface. Together with the two vacuum prescriptions, it fixes the full solution to the equation of motion of $\sigma$ based on both branches. This was discussed briefly in the flat-space example of ~\cite{GreenSun2025}. Schematically, the boundary saddle is
\begin{equation}
    \bar\sigma_*\sim iP_k\left[\int_{{\cal C}_+}d\tau\,a^4KJ_+-\int_{{\cal C}_-}d\tau\,a^4K^*J_-\right]\ .
\end{equation}
Alternatively, after plugging \eqref{eqn:simga_+_formal_sol} and \eqref{eqn:simga_-_formal_sol} back into the action, one obtains a gaussian integral in $\bar\sigma$ plus possible perturbative deviations. Evaluating the gaussian integral with proper perturbative expansions leads to results at generic order. We can re-sum the perturbative terms coming from a linear coupling of $\bar \sigma$ by completing the square. Substituting the full solution to $\bar \sigma$ back into $S_+-S_-$ then gives us the tree level result for integrating out the heavy field.

The resulting action depends only on the light fields. It contains nonlocal kernels, $G\sim(m^2-\Box)^{-1}$, and mixed-branch terms such as $J_+G_{+-}J_-$.\footnote{The functional integral does not require a large mass. However, for an authentic EFT treatment, one should do a large mass expansion to get around the non-locality. A heavy mass example is discussed in Section~\ref{subsec:qsfi-heavy-mass}.} These terms arise because the trace-fixed $\bar\sigma$ depends on the full contour, rather than on one branch alone. They are the non-factorizable terms in the reduced density matrix. In other words, having an effective action separately on each side of the in-in contour is insufficient for a complete effective description. This marks a fundamental distinction between effective theories for in-in correlators and standard EFTs for low energy scattering amplitudes.

One can nevertheless evaluate the time integrals and express $\bar\sigma_*$ in terms of $\bar\zeta_+$ and $\bar\zeta_-$. Its contribution can then be represented as a functional on the observation surface. It formally simplifies equal-time calculations, but does not actually factorize the reduced state. The boundary functional still depends on both branch variables, and only after explicitly performing the integral in time does it appear to depend only on boundary values of $\zeta$~\cite{GreenSun2025}.

In the following, we will follow these steps to work out examples for a quasi single field theory to see how the in-in correlators of the massless field can be recovered by a reduced density matrix.

\section{A quasi single field model}
\label{sec:qsfi}

In this section we study the general procedure in detail by analyzing a quasi-single-field model with a general light-field source. We will carry out the calculations of the reduced density matrix explicitly, and discuss the interpretations of the terms obtained.  The calculation below determines both contributions without an expansion in $m/H$, and then checks their sum against the original two field theory at the conformal mass. The heavy mass expansion is then discussed at the very end.

\subsection{The model and its sources}
\label{subsec:qsfi-model}

We work on the de Sitter background
\begin{align}
    ds^2={}&a^2(\tau)\left(-d\tau^2+d\mathbf{x}^2\right)\ ,
    \quad 
    a(\tau)=-\frac{1}{H\tau}\ ,
    \label{eq:qsfi-de-Sitter-metric}
\end{align}
with the conformal-coordinate Lagrangian
\begin{align}
    {\cal L} ={}& \frac{1}{2H^2\tau^2}\left[
    \zeta'^2-(\nabla\zeta)^2+\sigma'^2-(\nabla\sigma)^2
    -\frac{m^2}{H^2\tau^2}\sigma^2\right]
    +\frac{g}{H^2\tau^2}\,\sigma F[\zeta,\zeta']
    -\frac{\lambda}{3!H^4\tau^4}\sigma^3 \ .
    \label{eq:qsfi-original-action}
\end{align}
The field $\zeta$ denotes a canonically normalized massless scalar. A prime denotes a conformal-time derivative. The prescribed local source $F$ is independent of $\sigma$, may contain spatial derivatives, and obeys $F[0,0]=0$. The interaction is linear in $\sigma$, as in quasi-single-field constructions \cite{Chen2009we,Chen2009zp,NoumiYamaguchiYokoyama2013}, with an additional cubic self-interaction:
\begin{align}
    S[\zeta,\sigma] & = S_\zeta+S_\sigma+S_{\rm int} \ , \nonumber \\
    S_{\rm int} & =\int d\tau\,d^3x\left[a^2gF\sigma-\frac{a^4\lambda}{3!}\sigma^3\right]\ .
    \label{eq:qsfi-action-a}
\end{align}
We use constant couplings $g$ and $\lambda$. In particular, the linear mixing has measure $a^2g\sigma\zeta'$, whose time dependence differs from the often-used $a^3g\sigma\zeta'$ coupling. This choice affects the bispectrum and the local-limit phase cancellations below. We make this choice for better convergence of later calculations. Specifically, the bispectrum under our choice contains rational functions of the momenta, without special functions. See Appendix \ref{app:a3-mixing} for a detailed comparison with the $a^3 g \sigma\zeta'$ choice.

The two benchmark sources are\footnote{These two benchmarks stem from different models of quasi-single field inflation. For instance, $F_{\rm K}$ come from $\sigma(\partial T )^2$ where $T=\bar{T}(\tau) + \zeta(x)$ is the spontaneously broken clock field. }
\begin{equation}
    \begin{aligned}
        F_{\rm K}[\zeta] &\equiv \zeta'+c_{\rm K}\bigl[\zeta'^2-(\nabla\zeta)^2\bigr] \ , \\
        F_{\rm d}[\zeta] &\equiv \zeta'+3c_{\rm d}\zeta\zeta' = \partial_\tau\left( \zeta+\frac32c_{\rm d}\zeta^2 \right) \ .
    \end{aligned}
    \label{eq:qsfi-source-split}
\end{equation}
The source $F_{\rm d}$ gives a simple analytic bispectrum check, while $F_{\rm K}$ tests the kinetic derivative structure and its local limit. Their common linear term makes their order-$\lambda g^3$ bispectrum identical. Although $F_{\rm d}$ is a total derivative, the coupling $a^2\sigma F_{\rm d}$ is not a boundary term.

We take $a$ dimensionless and use canonical four-dimensional field dimensions,
\begin{align}
    [\zeta]=[\sigma]=[g]=[\lambda]=1\ , \qquad [F]=2\ , \qquad [c_{\rm d}]=-1\ , \qquad [c_{\rm K}]=-2\ .
    \label{eq:source-dimensions}
\end{align}
The constants $c_{\rm d}$ and $c_{\rm K}$ specify independent nonlinear-source scales, which will enter order-$g^2$ cubic kernels. The order-$\lambda g^3$ kernel, on the other hand, uses only the linear mixing. Superscripts on kernels count powers of $g$ and $\lambda$, with the source coefficients kept explicit.

The expansion in these couplings must remain perturbative over the momenta being compared. The quadratic mixing is of order $ga^2\zeta\sigma/\tau$. At $m\sim H$, its strength near horizon crossing is controlled by $|g|/k$. To have a well-controlled expansion, we then need
\begin{align}
    \frac{|g|}{k}\ll1 \ ,\qquad |c_{\rm d}|H\ll1 \ ,\qquad
    |c_{\rm K}|kH\ll1 \ ,\qquad \frac{|\lambda|}{H}\ll1 \ ,
    \label{eq:horizon-perturbative-estimates}
\end{align}
assuming the free scalar fluctuation amplitude is of order $H$ and applying the relevant source bound to each benchmark. More generally, the induced corrections to the light field power spectrum and to the massive field saddle must be small. Otherwise, one has to consider the other extreme of strong mixing~\cite{Huenupi:2026abj,Huenupi:2026aqc,Wang:2026lff,Pinol:2026xnl}. 

These models can be constructed from a spontaneously broken covariant model. For a clock $T(x)$ with background $T=\tau$, define
\begin{align}
    X_T \equiv -g^{\mu\nu}\nabla_\mu T\nabla_\nu T\ ,
    \qquad
    {\cal F}[\zeta;T] \equiv X_TF\!\left[\zeta , -\frac{g^{\mu\nu}\nabla_\mu T\nabla_\nu\zeta}{X_T} , -\frac{g^{\mu\nu} \nabla_\mu \zeta \nabla_\nu \zeta }{X_T} , \cdots \right]\ .
    \label{eq:qsfi-covariant-source}
\end{align}
The arguments of $F$ are scalars, as in the clock construction of the effective theory of inflation \cite{Cheung2007st}. With the mostly-plus convention, the covariant action is
\begin{align}
    S[\zeta,\sigma] ={}& \int d^4x\,\sqrt{-g}\,\left[
    -\frac12g^{\mu\nu}\nabla_\mu\zeta\nabla_\nu\zeta
    -\frac12g^{\mu\nu}\nabla_\mu\sigma\nabla_\nu\sigma
    -\frac{m^2}{2}\sigma^2+g\sigma{\cal F}[\zeta;T]
    -\frac{\lambda}{3!}\sigma^3\right]\ .
    \label{eq:qsfi-covariant-action}
\end{align}
On the chosen background, $X_T=a^{-2}$ and $-\nabla T\mathbin{\cdot}\nabla\zeta/X_T=\zeta'$, so ${\cal F}=a^{-2}F$. The kinetic source includes $-g^{\mu\nu}\nabla_\mu\zeta\nabla_\nu\zeta/X_T$ \cite{Achucarro2012sm}. 

At $\lambda=0$ the massive field integral is gaussian. We treat $\lambda$ perturbatively and work at tree level. The equation of motion is
\begin{align}
    {\cal D}_\sigma\sigma =gF-\frac{\lambda}{2}a^2\sigma^2\ ,
    \qquad
    {\cal D}_\sigma \equiv \partial_\tau^2-\frac{2}{\tau}\partial_\tau -\nabla^2+\frac{m^2}{H^2\tau^2}\ .
    \label{eq:qsfi-sigma-eom}
\end{align}
After integration by parts, the massive field action reads
\begin{align}
    S_\sigma+S_{\rm int} ={}& \int d\tau\,d^3x\left[-\frac{a^2}{2}\sigma{\cal D}_\sigma\sigma+ga^2F\sigma-\frac{\lambda a^4}{3!}\sigma^3\right] \nonumber\\
    &\quad
    +\frac12 \left[ \int d^3x\,a^2\sigma\sigma' \right]_{\tau_i}^{\tau_0} \ ,
    \label{eq:qsfi-sigma-action-ibp}
\end{align}
The final surface term is part of the fixed-boundary variational problem and must be included at finite $\tau_0$ \cite{GreenSun2025,Braglia2024Derivatives}. Our two operator conventions are related by
\begin{align}
    \Box-m^2=-a^{-2}{\cal D}_\sigma\ .
    \label{eq:qsfi-operator-sign-map}
\end{align}

\subsection{Exact non-local bulk effective action}
\label{subsec:qsfi-bulk-action}

Let $G_B(x,y)$ denote an inverse of ${\cal D}_\sigma$ with boundary conditions $B$, normalized by
\begin{align}
    {\cal D}_{\sigma,x}G_B(x,y) = \frac{\delta(\tau_x-\tau_y) \delta^3(\mathbf{x}-\mathbf{y})}{a^2(\tau_x)}\ .
    \label{eq:qsfi-green-normalization}
\end{align}
The label $B$ represents various kinds of boundary choices that we will encounter later. We now focus on the particular solution to the equation of motion for $\sigma$ and write
\begin{align}
    \sigma_{\rm cl} =\sigma^{(0)}+\lambda\sigma^{(1)} +{\cal O}(\lambda^2) \ .
    \label{eq:qsfi-lambda-solution-expansion}
\end{align}
The zeroth-order particular solution is
\begin{align}
    \sigma^{(0)}(x) =g\left({\cal D}_{\sigma,B}^{-1}F\right)(x)
    =  g\int d^4y\,a^2(\tau_y)G_B(x,y)F(y) \ .
    \label{eq:qsfi-sigma-classical}
\end{align}
At first order in $\lambda$,
\begin{align}
    {\cal D}_\sigma\sigma^{(1)} ={}&-\frac12a^2\left(\sigma^{(0)}\right)^2, \nonumber\\
    \sigma^{(1)}(x) ={}&-\frac12 \int d^4y\,a^4(\tau_y)G_B(x,y) \left[\sigma^{(0)}(y)\right]^2 \ .
    \label{eq:qsfi-lambda-sigma-one}
\end{align}
The correction $\sigma^{(1)}$ is necessary for reconstructing the heavy field at this order, but since the action is stationary on $\sigma^{(0)}$ at $\lambda = 0$, the terms linear in $\sigma^{(1)}$ cancel in the action. Substituting the saddle into the bulk action therefore gives
\begin{align}
    S_{\rm eff}^{\rm bulk}[\zeta] ={}& S_\zeta
    +\frac{g^2}{2}\int d^4x\,d^4y\,a^2(\tau_x)a^2(\tau_y)F(x)G_B(x,y)F(y)\nonumber\\
    &-\frac{\lambda g^3}{3!}\int d^4x\,a^4(\tau_x)
    \left[\int d^4y\,a^2(\tau_y)G_B(x,y)F(y)\right]^3+{\cal O}(\lambda^2) \ .
    \label{eq:qsfi-exact-effective-action}
\end{align}
This is the usual in-out EFT derivation.

At $\lambda=0$, the quadratic operator of $\sigma$, ${\cal D}_\sigma$, does not depend on $\zeta$, so the gaussian functional determinant is $\zeta$ independent and changes only the normalization.  When $\lambda\neq0$, the Hessian about the saddle contains $\lambda a^2\sigma^{(0)}$ and is source dependent. Its determinant is a one-loop effect, not part of the tree-level action in \eqref{eq:qsfi-exact-effective-action}. At finite time, the boundary fluctuation determinant belongs to the same loop calculation; both determinants are outside the tree-level approximation used here.

For either source in \eqref{eq:qsfi-source-split}, the term quadratic in $F$ generates light-field interactions of orders two through four, whereas the term cubic in $F$ generates orders three through six. We keep $F$ general until Section~\ref{subsec:qsfi-probability-cubic-kernel}, where the terms relevant to the bispectrum are selected.

\subsection{The finite-time wavefunctional and influence functional}
\label{subsec:qsfi-finite-time-influence}

The boundary path integral over $\bar{\sigma}$ turns the fixed boundary bulk action into an influence action as it generates the mixed branch contractions. We first derive this for the gaussian sector and then turn on the effect of $\sigma^3$. Write $F_s\equiv F[\zeta_s,\zeta_s']$ on branch $s=\pm$. The trace identifies the final massive field values,
\begin{align}
    \sigma_+(\tau_0,\mathbf{x})=\sigma_-(\tau_0,\mathbf{x})=\bar\sigma(\mathbf{x})\ .
    \label{eq:qsfi-shared-heavy-boundary-value}
\end{align}
It is useful to keep the branch-dependent Bunch--Davies contours explicit:
\begin{align}
    {\cal C}_+:\quad\tau_{i,+}=-\infty(1-i\epsilon)\longrightarrow\tau_0 \ ,
    \qquad
    {\cal C}_-:\quad\tau_{i,-}=-\infty(1+i\epsilon)\longrightarrow\tau_0 \ .
    \label{eq:qsfi-branch-contours}
\end{align}

Let $u_k(\tau)$ be the Bunch--Davies mode for $\sigma$ on the $+$ branch with the Wronskian
\begin{align}
    a^2(\tau) \left[u_k(\tau)u_k^{*\prime}(\tau)-u_k'(\tau)u_k^*(\tau)\right]=-i \ .
    \label{eq:qsfi-heavy-wronskian}
\end{align}
Define
\begin{align}
    K_{+,k}(\tau) ={}&\frac{u_k(\tau)}{u_k(\tau_0)} \ ,
    \qquad
    P_k(\tau_0)=|u_k(\tau_0)|^2 \ , \nonumber\\
    \Omega_{+,k} ={}&a_0^2K_{+,k}'(\tau_0)\ ,
    \qquad
    a_0 = a(\tau_0)\ ,
    \label{eq:qsfi-boundary-kernels}
\end{align}
and, on the anti-time-ordered branch,
\begin{align}
    K_{-,k}=K_{+,k}^*\ ,
    \qquad
    \Omega_{-,k}=\Omega_{+,k}^*\ .
    \label{eq:qsfi-minus-kernels}
\end{align}
The Schwinger--Keldysh propagators in these conventions are 
\begin{align}
    G_{+- , k}(\tau,\tau') =  iP_k K_{+,k}(\tau) K_{-,k}(\tau') \ , \quad G_{-+,k}(\tau,\tau') = iP_k K_{-,k}(\tau) K_{+,k}(\tau') \ , \\
    G_{++,k}(\tau,\tau') = G_{+-,k}(\tau,\tau')\theta(\tau' - \tau) + G_{-+,k}(\tau,\tau') \theta(\tau-\tau') \ , \quad G_{--,k} = G_{++,k}^* \ .
\end{align}
From the Wronskian one sees
\begin{align}
    \Delta \Omega_k \equiv \Omega_{+,k}-\Omega_{-,k} ={}&a_0^2\left[\frac{u_k'(\tau_0)}{u_k(\tau_0)}-\frac{u_k^{*\prime}(\tau_0)}{u_k^*(\tau_0)}\right]=\frac{i}{P_k} \ , \nonumber\\
    \operatorname{Im}\Omega_{+,k}={}&\frac{1}{2P_k}>0 \ .
    \label{eq:qsfi-Omega-P-relation}
\end{align}
The inequality ensures normalizability for the gaussian heavy field wavefunctional.

Let $G_{D,\pm,k}$ be the ``$\pm$'' branch propagator with a homogeneous Dirichlet condition at the final time, we have
\begin{align}
    G_{D,\pm,k}(\tau_0,\tau') = {}&G_{D,\pm,k}(\tau,\tau_0)=0 \ , \\
    G_{D,\pm,k}(\tau, \tau') = {}& G_{\pm\pm,k}(\tau,\tau') \mp i P_k K_{\pm,k}(\tau) K_{\pm}(\tau') \ ,
    \label{eq:qsfi-dirichlet-green}
\end{align}
This is the bulk-to-bulk propagator used in the wavefunction literature. The corresponding Green identity gives
\begin{align}
    a_0^2 \partial_\tau G_{D,+,k}(\tau,\tau') \big|_{\tau=\tau_0}=K_{+,k}(\tau') \ .
    \label{eq:qsfi-green-boundary-identity}
\end{align}
Define the source integrals on the two branches,
\begin{align}
    I_{\pm,\mathbf{k}}[F_\pm] \equiv{}& \int_{{\cal C}_\pm}d\tau\, a^2(\tau)K_{\pm,k}(\tau)F_{\pm,\mathbf{k}}(\tau) \ .
    \label{eq:qsfi-I-pm}
\end{align}
The two functionals are independent: $I_-=I_+^*$ only after the two real histories are identified.

We are now ready to calculate the fixed-boundary on-shell action. At fixed $\sigma_{\pm,\mathbf{k}}(\tau_0)=\bar\sigma_{\mathbf{k}}$, the zeroth-order solution is
\begin{align}
    \sigma^{(0)}_{\pm,\mathbf{k}}(\tau) = \bar\sigma_{\mathbf{k}}K_{\pm,k}(\tau)+g\int_{{\cal C}_\pm}d\tau'\,a^2(\tau')G_{D,\pm,k}(\tau,\tau')F_{\pm,\mathbf{k}}(\tau') \ .
    \label{eq:qsfi-fixed-boundary-solution}
\end{align}
Plugging back into \eqref{eq:qsfi-sigma-action-ibp}, together with equation \eqref{eq:qsfi-green-boundary-identity}, we have the zeroth-order-in-$\lambda$ result 
\begin{align}
    S_{\sigma,\pm}^{(0),{\rm os}} ={}& \frac12 \int_{\mathbf{k}} \Omega_{\pm,k}\bar\sigma_{\mathbf{k}}\bar\sigma_{-\mathbf{k}} +g\int_{\mathbf{k}} \bar\sigma_{-\mathbf{k}}I_{\pm,\mathbf{k}} \nonumber\\
    &+\frac{g^2}{2}\int_{\mathbf{k}}\int_{{\cal C}_\pm}d\tau\int_{{\cal C}_\pm}d\tau'\,a^2(\tau)a^2(\tau')F_{\pm,\mathbf{k}}(\tau)G_{D,\pm,k}(\tau,\tau')F_{\pm,-\mathbf{k}}(\tau') \ .
    \label{eq:qsfi-fixed-boundary-onshell-action}
\end{align}
We use the shorthand $\int_{\mathbf{k}}\equiv\int\frac{d^3k}{(2\pi)^3} $.  At first order in $\lambda$, the fixed-boundary on-shell action is 
\begin{align}
    S_{\sigma,\pm}^{\rm os}[\bar\sigma,F_\pm]=S_{\sigma,\pm}^{(0),{\rm os}}[\bar\sigma,F_\pm]-\frac{\lambda}{3!}\int_{{\cal C}_\pm}d^4x\,a^4(\tau)\left[\sigma_\pm^{(0)}(x)\right]^3+{\cal O}(\lambda^2)\ .
    \label{eq:qsfi-fixed-boundary-lambda-onshell}
\end{align}
Compare with \eqref{eq:qsfi-exact-effective-action}, we gain additional terms involving $\bar\sigma$. 

\begin{figure}[t]
\centering
\begin{tikzpicture}[x=1cm,y=1cm,>=stealth,font=\small]
    \draw[->,thick] (0,0.45) -- (4,0.05);
    \draw[->,thick] (4,-0.05) -- (0,-0.45);
    \draw[dashed] (4,0.7) -- (4,-0.7);
    \node[above] at (2,0.45) {$\sigma_+$};
    \node[below] at (2,-0.45) {$\sigma_-$};
    \node[left] at (0,0.45) {BD};
    \node[left] at (0,-0.45) {BD$^*$};
    \node[right] at (4,0) {$\bar\sigma$};
    \node[below] at (4,-0.8) {$\tau_0$};
    \draw[->,thick] (5.2,0) -- (7,0);
    \node[above] at (6.1,0.1) {$\displaystyle\int{\cal D}\bar\sigma$};
    \node[anchor=west,align=left] at (7.35,0.45)
      {same branch: $G_D\ \longrightarrow\ G_F$~\eqref{eq:qsfi-Feynman-kernel}};
    \node[anchor=west,align=left] at (7.35,-0.45)
      {mixed branches: $G_{+-},\ G_{-+}$~\eqref{eq:qsfi-gaussian-SK-form}};
\end{tikzpicture}
\caption{The massive field trace at the observation time. Integrating over the common final value supplies both the completion of the fixed-boundary propagators and the mixed-branch contractions. The two final light-field values remain independent in the reduced density matrix. }
\label{fig:finite-time-trace}
\end{figure}
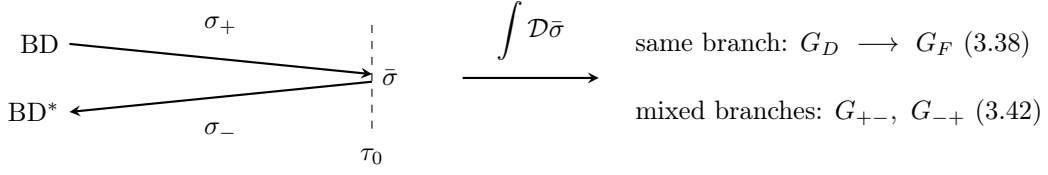

Taking the trace over $\bar\sigma$ in the path integral gives the reduced density matrix
\begin{align}
    \rho_{\rm red}[\bar\zeta_+,\bar\zeta_-,\tau_0] &=\int{\cal D}\bar\sigma\,\Psi[\bar\zeta_+,\bar\sigma]\Psi^*[\bar\zeta_-,\bar\sigma] \\
    & =\int_{\substack{\zeta_+(\tau_0)=\bar\zeta_+\\\zeta_-(\tau_0)=\bar\zeta_-}}
    {\cal D}\zeta_+{\cal D}\zeta_-\,
    e^{iS_\zeta[\zeta_+]-iS_\zeta[\zeta_-]+iS_{\rm IF}[\zeta_+,\zeta_-]} \ .
    \label{eq:qsfi-reduced-density-matrix}
\end{align}
The initial state is fixed by the branch prescriptions. We calculate the influence action first, and evaluate the light field on its tree-level saddles in Section~\ref{subsec:qsfi-probability-cubic-kernel}. Introduce
\begin{align}
    \Delta I_{\mathbf{k}} \equiv I_{+,\mathbf{k}}-I_{-,\mathbf{k}}\ .
    \label{eq:qsfi-boundary-deltas}
\end{align}
The terms in the exponent $iS_{\rm SK}=iS_+-iS_-$ that depend on $\bar\sigma$ are
\begin{align}
    iS_{\rm SK} \supset \frac{i}{2}\int_{\mathbf{k}}\Delta\Omega_k\,\bar\sigma_{\mathbf{k}}\bar\sigma_{-\mathbf{k}}+ig\int_{\mathbf{k}}\bar\sigma_{-\mathbf{k}}\Delta I_{\mathbf{k}} \ .
    \label{eq:qsfi-boundary-gaussian-exponent}
\end{align}
As $\Delta\Omega_k=i/P_k$, this gaussian exponent is evaluated at the saddle
\begin{align}
    \bar\sigma_{*,\mathbf{k}} ={}& -g\frac{\Delta I_{\mathbf{k}}}{\Delta\Omega_k} =igP_k\Delta I_{\mathbf{k}}\ .
    \label{eq:qsfi-boundary-saddle}
\end{align}
This is the common boundary value selected by the condition \eqref{eq:general-momentum-sewing}. Note that it is of order $g$. The sources on the two branches need not coincide, so the saddle is generally complex. At ${\cal O}(\lambda^0)$ the integral gives
\begin{align}
    iS_{{\rm SK},{\rm EFT}}\supset -\frac{g^2}{2}\int_{\mathbf{k}}P_k\Delta I_{\mathbf{k}}\Delta I_{-\mathbf{k}}\ .
    \label{eq:qsfi-boundary-gaussian-result}
\end{align}
Although generated by the final time trace, this term depends on the bulk histories through $I_{\mathbf k}$.  Restoring those integrals makes its nonlocal and mixed-branch content explicit \cite{GreenSun2025,Cespedes2025Quartic}. At $\lambda=0$ the gaussian determinant is source independent and cancels in normalization. At nonzero $\lambda$, the boundary Hessian can depend on the light source, and its determinant belongs to a one-loop calculation together with the bulk determinant\cite{Cespedes2023IR}. We will stay in tree level in the rest of the work. Figure~\ref{fig:finite-time-trace} summarizes the contour construction.

The connection to a single branch bulk EFT calculation is visible upon switching the Dirichlet propagator to the Feynman propagator, 
\begin{align}
    G_{F,k}(\tau,\tau') = G_{++,k}(\tau,\tau') = {}& G_{D,+,k}(\tau,\tau') + iP_k K_{+,k}(\tau) K_{+,k}(\tau')\ .
    \label{eq:qsfi-Feynman-kernel}
\end{align}
The same-branch Dirichlet contribution can thus be written as
\begin{align}
    &\frac{g^2}{2} \int_{\mathbf{k}} \int_{{\cal C}_\pm}d\tau\int_{{\cal C}_\pm}d\tau'\,a^2(\tau)a^2(\tau')F_{\pm,\mathbf{k}}(\tau)G_{D,\pm,k}(\tau,\tau')F_{\pm,-\mathbf{k}}(\tau')\nonumber\\
    =&\frac{g^2}{2}\int_{\mathbf{k}}\int_{{\cal C}_\pm}d\tau\int_{{\cal C}_\pm}d\tau'\,a^2(\tau)a^2(\tau')F_{\pm,\mathbf{k}}(\tau)G_{\pm\pm,k}(\tau,\tau')F_{\pm,-\mathbf{k}}(\tau') \mp \frac{ig^2}{2}\int_{\mathbf{k}}P_kI_{\pm,\mathbf{k}}I_{\pm,-\mathbf{k}} \ .
    \label{eq:qsfi-plus-Dirichlet-Feynman}
\end{align}
On the other hand, 
\begin{align}
    -\frac{g^2}{2}\int_{\mathbf{k}}P_k \Delta I_{\mathbf{k}}\Delta I_{-\mathbf{k}} 
    ={}-\frac{g^2}{2}\int_{\mathbf{k}}P_k\left(I_{+,\mathbf{k}}I_{+,-\mathbf{k}}+ I_{-,\mathbf{k}}I_{-,-\mathbf{k}}\right) + g^2\int_{\mathbf{k}}P_kI_{+,\mathbf{k}}I_{-,-\mathbf{k}} \ ,
    \label{eq:qsfi-boundary-exponent-expanded}
\end{align}
where we used $P_{-k}=P_k$ and relabeled $\mathbf{k}\rightarrow-\mathbf{k}$ in one of the two mixed terms. 

Combining the bulk and trace contributions gives the quadratic influence action,
\begin{align}
  iS_{\rm IF}^{(2)}    & = iS^{(0),\text{os}}_{\sigma,+} - iS^{(0),\text{os}}_{\sigma,-} \\&=  \frac{ig^2}{2} \int_{\mathbf k} \bigg( \int_{{\cal C}_+}d\tau\int_{{\cal C}_+}d\tau'\,a^2(\tau)a^2(\tau')F_{+,\mathbf{k}}(\tau)G_{F,k}(\tau,\tau')F_{+,-\mathbf{k}}(\tau') \nonumber\\
    & \qquad \ \  - \int_{{\cal C}_-}d\tau\int_{{\cal C}_-}d\tau'\,a^2(\tau)a^2(\tau')F_{-,\mathbf{k}}(\tau)G_{F,k}^*(\tau,\tau')F_{-,-\mathbf{k}}(\tau') \bigg) \nonumber \\
    & \qquad +g^2\int_{\mathbf{k}}P_kI_{+,\mathbf{k}}I_{-,-\mathbf{k}} \ .
    \label{eq:qsfi-gaussian-SK-form}
\end{align}
The last line is the mixed-branch contribution that cannot be specified by an in-out bulk action alone. The sum agrees with the standard in-in propagator organization \cite{Chen2017ryl,Cespedes2025Quartic,Palma2026Map}.

The same boundary saddle also fixes the order-$\lambda$ contribution. Plugging $\bar\sigma_*$ back to the zeroth order solution \eqref{eq:qsfi-fixed-boundary-solution}, we have 
\begin{equation}
      \sigma_{\pm,\mathbf{k}}^{(0)} = g \int_{\cal{C}_\pm} d\tau' \ a^2 (\tau') G_{\pm\pm}(\tau,\tau') F_{\pm,\mathbf{k}}(\tau') \mp g \int_{\cal C_\mp} d\tau' \ a^2(\tau') G_{\pm\mp,k} (\tau,\tau') F_{\mp,\mathbf{k}}(\tau') \ .
    \label{eq:qsfi-sigma-pm-SK}
\end{equation} 
Note that it contains contributions from both branches. The tree-level order-$\lambda$ term in the exponent is
\begin{equation}
    iS_{\rm IF}^{(\lambda)} = -\frac{i\lambda }{3!} \left(\int_{{\cal C}_+}d^4x\,a^4 (\sigma_{+}^{(0)}(x))^3 - \int_{{\cal C}_-}d^4x\,a^4(\sigma_{-}^{(0)}(x))^3\right)\ .
    \label{eq:qsfi-lambda-Feynman-Wightman-compact}
\end{equation}
Here $\sigma_\pm^{(0)}(x)$ denotes the position-space form of \eqref{eq:qsfi-sigma-pm-SK}. We will see explicit results of this in a concrete calculation later. 

Putting the gaussian terms and the tree-level order-$\lambda$ terms together, with every quadratic convolution displayed explicitly,
\begin{align}
    \rho_{\rm red}[\bar\zeta_+,\bar\zeta_-] \propto{}& \int_{\zeta_\pm(\tau_0)=\bar\zeta_\pm}
    {\cal D}\zeta_+{\cal D}\zeta_-\, \exp\Bigg\{ iS_\zeta[\zeta_+]-iS_\zeta[\zeta_-] \nonumber\\
    &\quad +\frac{ig^2}{2} \int_{\mathbf{k}} \int_{{\cal C}_+}d\tau \int_{{\cal C}_+}d\tau'\, a^2(\tau)a^2(\tau') F_{+,\mathbf{k}}(\tau) G_{F,k}(\tau,\tau') F_{+,-\mathbf{k}}(\tau') \nonumber\\
    &\quad -\frac{ig^2}{2} \int_{\mathbf{k}} \int_{{\cal C}_-}d\tau \int_{{\cal C}_-}d\tau'\, a^2(\tau)a^2(\tau') F_{-,\mathbf{k}}(\tau) G_{F,k}^*(\tau,\tau') F_{-,-\mathbf{k}}(\tau') \nonumber\\
    &\quad + g^2\int_{\mathbf{k}} P_kI_{+,\mathbf{k}}I_{-,-\mathbf{k}}   +iS_{\rm IF}^{(\lambda)} +{\cal O}(\lambda^2)  \Bigg\} \ .
    \label{eq:qsfi-reduced-density-matrix-final}
\end{align}

The explicit cross-branch convolution at this order is
\begin{align}
    g^2\int_{\mathbf{k}} P_kI_{+,\mathbf{k}}I_{-,-\mathbf{k}} ={}& g^2\int_{\mathbf{k}} \int_{{\cal C}_+}d\tau \int_{{\cal C}_-}d\tau'\, a^2(\tau)a^2(\tau')P_k \nonumber\\
    &\qquad\times K_{+,k}(\tau)K_{-,k}(\tau') F_{+,\mathbf{k}}(\tau) F_{-,-\mathbf{k}}(\tau') \ .
    \label{eq:qsfi-cross-branch-convolution}
\end{align}
This term mixes the $+$ and $-$ histories and therefore cannot be written as $S[\zeta_+]-S[\zeta_-]$. It is a genuine influence term in the reduced density matrix. At order $\lambda$, since $\sigma_{+}^{(0)}(x)$ contains convolutions both in $+$ and in $-$ branches, its cube introduces mixing terms that cannot be separated into two branches. The propagators involved in \eqref{eq:qsfi-sigma-pm-SK} satisfies $G_F-G_{+-}=G_F^*+G_{-+}=G_R$, so that when $F_+=F_-$,
\begin{equation}
    \sigma_+^{(0)} = \sigma_-^{(0)} = g \int d\tau' \ a^2(\tau') G_R(\tau,\tau')F(\tau') \ .
\end{equation}
The two terms in \eqref{eq:qsfi-lambda-Feynman-Wightman-compact} then cancel after this identification\footnote{One might be worried that here we are mixing the two contours with different vacuum prescriptions. To get around this, calculate $\sigma_+^{(0)} - \sigma_-^{(0)}$, collect integrals on $\cal{C}_+$ and $\cal{C}_-$ respectively. One sees clearly that when $\tau>\tau'$, the propagators cancel within each contour. When $\tau'>\tau$, we are free to put both contours on the real axis to obtain the final cancellation. }. Thus the complete tree-level influence functional is normalized when the histories are identified. The appearance of $G_R$ is also not surprising as it encodes the classical response from the source. 

\subsection{Cubic results}
\label{subsec:qsfi-probability-cubic-kernel}

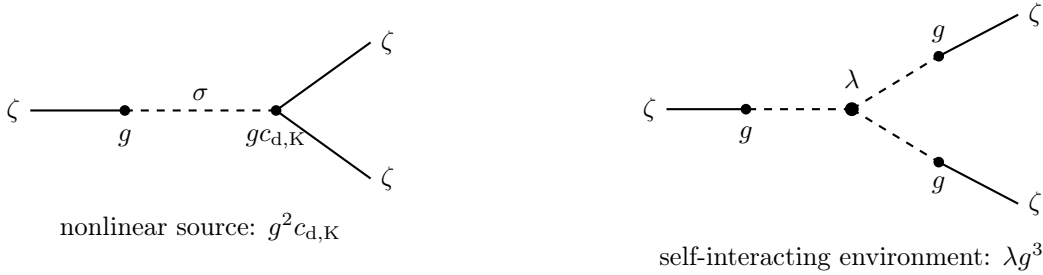
\begin{figure}[h]
\centering
\begin{minipage}{0.47\textwidth}
\centering
\begin{tikzpicture}[x=1cm,y=1cm,font=\small]
    \draw[thick] (0,0) -- (1.25,0);
    \draw[thick,dashed] (1.25,0) -- (3.25,0);
    \draw[thick] (3.25,0) -- (4.5,0.9);
    \draw[thick] (3.25,0) -- (4.5,-0.9);
    \fill (1.25,0) circle (2pt);
    \fill (3.25,0) circle (2pt);
    \node[above] at (2.25,0) {$\sigma$};
    \node[below] at (1.25,-0.12) {$g$};
    \node[below] at (3.25,-0.12) {$g c_{\rm d,K}$};
    \node[left] at (0,0) {$\zeta$};
    \node[right] at (4.5,0.9) {$\zeta$};
    \node[right] at (4.5,-0.9) {$\zeta$};
    \node at (2.25,-1.55) {nonlinear source: $g^2c_{\rm d,K}$};
\end{tikzpicture}
\end{minipage}\hfill
\begin{minipage}{0.50\textwidth}
\centering
\begin{tikzpicture}[x=1cm,y=1cm,font=\small]
    \draw[thick,dashed] (2.1,0) -- (0.7,0);
    \draw[thick,dashed] (2.1,0) -- (3.25,0.7);
    \draw[thick,dashed] (2.1,0) -- (3.25,-0.7);
    \draw[thick] (0.7,0) -- (-0.35,0);
    \draw[thick] (3.25,0.7) -- (4.3,1.25);
    \draw[thick] (3.25,-0.7) -- (4.3,-1.25);
    \fill (2.1,0) circle (2.6pt);
    \fill (0.7,0) circle (2pt);
    \fill (3.25,0.7) circle (2pt);
    \fill (3.25,-0.7) circle (2pt);
    \node[above] at (2.1,0.15) {$\lambda$};
    \node[below] at (0.7,-0.1) {$g$};
    \node[above] at (3.25,0.75) {$g$};
    \node[below] at (3.25,-0.75) {$g$};
    \node[left] at (-0.35,0) {$\zeta$};
    \node[right] at (4.3,1.25) {$\zeta$};
    \node[right] at (4.3,-1.25) {$\zeta$};
    \node at (2.1,-1.95) {self-interacting environment: $\lambda g^3$};
\end{tikzpicture}
\end{minipage}
\caption{Two contributions to the $\zeta$ bispectrum. Solid lines denote $\zeta$ legs and dashed lines denote $\sigma$ propagation. A linear and a quadratic source insertion give the order $g^2c_{\rm d,K}$ contribution through a gaussian influence action; $c_{\rm d,K}$ denotes the coefficient of the chosen source. Three linear mixings with a $\sigma^3$ vertex give the order-$\lambda g^3$ contribution. The derivative factors and contractions are given in the text.}
\label{fig:bispectrum-mechanisms}
\end{figure}

There are two tree-level diagrams for the light field bispectrum. The term quadratic in $F$ generates an order-$g^2$ cubic light interaction because $F$ is nonlinear in $\zeta$. The massive field self interaction instead generates an order-$\lambda g^3$ contribution through an influence action cubic in $F$. Figure~\ref{fig:bispectrum-mechanisms} shows the two mechanisms.

We now evaluate the cubic terms in \eqref{eq:qsfi-reduced-density-matrix-final}. To distinguish the light and heavy propagators, we write $K^\sigma_{\pm,k}\equiv K_{\pm,k}$ and $G^\sigma_{ss',k}\equiv G_{ss',k}$. The light field is evaluated on its free bulk-to-boundary saddle,
\begin{align}
    \zeta_{s,\mathbf k}(\tau)
    =K^\zeta_{s,k}(\tau)\bar\zeta_{s,\mathbf k}\ ,
    \qquad
    K^\zeta_{-,k}=(K^\zeta_{+,k})^*\ ,
    \qquad
    K^\zeta_{+,k}(\tau)
    =\frac{(1-ik\tau)e^{ik\tau}}
    {(1-ik\tau_0)e^{ik\tau_0}} \ .
    \label{eq:probability-light-branch-saddles}
\end{align}

Define the branch-resolved cubic kernels by
\begin{align}
    \left.\ln\rho_{\rm red}\right|_{\bar\zeta^3}
    =-\frac{1}{3!}
    \int_{\mathbf k_1\mathbf k_2\mathbf k_3}
    (2\pi)^3\delta^3\!\left(\sum_{a=1}^3\mathbf k_a\right)
    \sum_{s_1,s_2,s_3=\pm}
    \Gamma^{(3)}_{s_1s_2s_3}(1,2,3)
    \prod_{a=1}^3\bar\zeta_{s_a,\mathbf k_a}\ .
    \label{eq:probability-branch-cubic-definition}
\end{align}
On the diagonal, $\tilde\rho_\zeta[\bar\zeta]\equiv \rho_{\rm red}[\bar\zeta,\bar\zeta]$, write
\begin{align}
    \left.\ln\tilde\rho_\zeta\right|_{\bar\zeta^3}
    =-\frac{1}{3!}
    \int_{\mathbf k_1\mathbf k_2\mathbf k_3}
    (2\pi)^3\delta^3\!\left(\sum_a\mathbf k_a\right)
    \Gamma^{(3)}_{\tilde\rho}(1,2,3)
    \prod_a\bar\zeta_{\mathbf k_a} \ .
    \label{eq:probability-diagonal-cubic-definition}
\end{align}
Hermiticity and simultaneous permutation of momentum and branch labels imply
\begin{align}
    \Gamma^{(3)}_{---}=\left(\Gamma^{(3)}_{+++}\right)^* \ ,
    \qquad
    \Gamma^{(3)}_{+--}(i;j,\ell)
    =\left(\Gamma^{(3)}_{++-}(j,\ell;i)\right)^* \ .
    \label{eq:probability-hermiticity-cubic}
\end{align}
Therefore the complete diagonal kernel only requires the structure
\begin{align}
    \Gamma^{(3)}_{\tilde\rho}(1,2,3) = 2\,\mathrm{Re}\!\left[\Gamma^{(3)}_{+++}(1,2,3) +\sum_{{\rm cyc}(i,j,\ell)} \Gamma^{(3)}_{++-}(j,\ell;i) \right] \ .
    \label{eq:probability-diagonal-branch-sum}
\end{align}
Summing the branches converts the matched influence data into the field probability. The leading order bispectrum is then
\begin{align}
    \left\langle \bar\zeta_{\mathbf k_1}\bar\zeta_{\mathbf k_2}\bar\zeta_{\mathbf k_3} \right\rangle_c' =-\Gamma^{(3)}_{\tilde\rho}(1,2,3) \prod_{a=1}^3P_\zeta(k_a)+\cdots \ .
    \label{eq:probability-cubic-kernel-bispectrum}
\end{align}
Here $P_\zeta(k)$ is the free equal-time light-field power spectrum, which approaches $H^2/(2k^3)$ as $\tau_0\to0$. The prime removes the momentum-conserving delta function.

We start with order $g^2$ . For $F_{\rm d}=\zeta'+3c_{\rm d}\zeta\zeta'$, direct differentiation of \eqref{eq:qsfi-gaussian-SK-form} gives
\begin{align}
    \Gamma^{(3,g^2;{\rm d})}_{+++}(1,2,3) &=-3i c_{\rm d} g^2\sum_{{\rm cyc}(i,j,\ell)} \int_{{\cal C}_+}d\tau d\tau'\, a^2(\tau)a^2(\tau')\nonumber\\
    &\times K^{\zeta\prime}_{+,k_i}(\tau) G^\sigma_{++,k_i}(\tau,\tau') \partial_{\tau'}\!\left[ K^\zeta_{+,k_j}(\tau')K^\zeta_{+,k_\ell}(\tau') \right] \ ,
    \label{eq:probability-Fd-+++} \\
    \Gamma^{(3,g^2;{\rm d})}_{++-}(1,2;3) &=3i c_{\rm d} g^2 \int_{{\cal C}_+}d\tau\int_{{\cal C}_-}d\tau'\, a^2(\tau)a^2(\tau')\nonumber\\
    &\times \partial_\tau\!\left[ K^\zeta_{+,k_1}(\tau)K^\zeta_{+,k_2}(\tau) \right] G^\sigma_{+-,k_3}(\tau,\tau') K^{\zeta\prime}_{-,k_3}(\tau') \ .
    \label{eq:probability-Fd-++-}
\end{align}
Together with \eqref{eq:probability-hermiticity-cubic}, these two equations give the complete order-$g^2$ matching for $F_{\rm d}$. For the kinetic source $F_{\rm K}=\zeta'+c_{\rm K}\big[\zeta'^2-(\nabla\zeta)^2\big] $, one replaces the quadratic vertex above by
\begin{align}
    3c_{\rm d}\,\partial_\tau(K^\zeta_{s,k_j}K^\zeta_{s,k_\ell})
    \quad\longrightarrow\quad
    2c_{\rm K}\left[ K^{\zeta\prime}_{s,k_j}K^{\zeta\prime}_{s,k_\ell} +\mathbf k_j\!\cdot\!\mathbf k_\ell K^\zeta_{s,k_j}K^\zeta_{s,k_\ell} \right],
    \label{eq:probability-FK-replacement}
\end{align}
with $\mathbf k_j\!\cdot\!\mathbf k_\ell =(k_i^2-k_j^2-k_\ell^2)/2$ in the channel where the massive line carries $k_i$. The explicit $F_{\rm K}$ branch formulas are collected in Appendix~\ref{app:kinetic-kernels}.

Now move to order $\lambda g^3$. Inside the $\sigma^3$ term only the linear term in source $F_s=\zeta_s'$ contributes. Using \eqref{eq:qsfi-sigma-pm-SK}, the complete diagonal contribution is
\begin{align}
    \begin{aligned}
        \Gamma^{(3,\lambda)}_{\tilde\rho}(1,2,3) ={}&-2\lambda g^3\,\mathrm{Im}  \int_{{\cal C}_+}d\tau\,a^4(\tau) \prod_{i=1}^3 \Bigg[ \int_{{\cal C}_+}d\tau_i\,a^2(\tau_i) G^\sigma_{++,k_i}(\tau,\tau_i) K^{\zeta\prime}_{+,k_i}(\tau_i) \\[-1mm]
        &\hspace{35mm} -\int_{{\cal C}_-}d\tau_i\,a^2(\tau_i) G^\sigma_{+-,k_i}(\tau,\tau_i) K^{\zeta\prime}_{-,k_i}(\tau_i) \Bigg] \ .
    \end{aligned}
    \label{eq:probability-lambda-diagonal-kernel}
\end{align}
This term is common to $F_{\rm d}$ and $F_{\rm K}$ because their linear mixing terms are the same. The full cubic kernel entering the bispectrum is then
\begin{align}
    \Gamma^{(3)}_{\tilde\rho} =\Gamma^{(3,g^2)}_{\tilde\rho} +\Gamma^{(3,\lambda)}_{\tilde\rho} +{\cal O}(g^4,\lambda g^5,\lambda^2) \ .
    \label{eq:probability-total-cubic-kernel}
\end{align}

\subsection{Conformally coupled benchmark} 
A useful example in which all time integrals can be done in elementary form is the conformally coupled case $m^2=2H^2$, or $\nu=1/2$. Taking $\tau_0\to0$ after forming the Schwinger--Keldysh propagators, and defining $k_t=k_1+k_2+k_3$, the complete order-$g^2$ result for $F_{\rm d}$ is
\begin{align}
    \Gamma^{(3,g^2;{\rm d})}_{\tilde\rho}(1,2,3) = -\frac{6 c_{\rm d} g^2}{H^2} \left[k_t -\frac{k_1k_2+k_2k_3+k_3k_1}{k_t} -\frac{k_1k_2k_3}{k_t^2} \right] \ .
    \label{eq:probability-conformal-g2}
\end{align}
Similarly, the order $\lambda g^3$ result is
\begin{align}
    \Gamma^{(3,\lambda)}_{\tilde\rho}(1,2,3) = -\frac{\pi\lambda g^3}{8H^4}\ .
    \label{eq:probability-conformal-lambda}
\end{align}
Thus, for this benchmark, 
\begin{align}
    \left\langle\bar\zeta_{\mathbf k_1} \bar\zeta_{\mathbf k_2}\bar\zeta_{\mathbf k_3}\right\rangle_c' = \left\{ \frac{6 c_{\rm d} g^2}{H^2} \left[ k_t-\frac{k_1k_2+k_2k_3+k_3k_1}{k_t} -\frac{k_1k_2k_3}{k_t^2} \right] +\frac{\pi\lambda g^3}{8H^4} \right\} \prod_{a=1}^3P_\zeta(k_a)+\cdots \ .
    \label{eq:probability-conformal-bispectrum}
\end{align}
Appendix~\ref{app:conformal-check} derives these expressions from the reduced description and from the original two-field Hamiltonian. The latter includes the Legendre transform term required by the derivative coupling and reproduces both contributions in \eqref{eq:probability-conformal-bispectrum}. 

In \cite{GreenSun2025} it is shown that corrections from the homogeneous part of the solution to the equation of motion redshift away at $\tau \to 0$ at perturbative order in $H/m$, for a heavy mass mediator with mass $m$. We will see this behavior in the heavy mass limit in section \ref{subsec:qsfi-heavy-mass}. Here for conformally coupled mediator, however, we cannot simply drop the homogeneous solution. For example, setting $\bar\sigma = 0$, the value $\Gamma_{\tilde \rho}^{(3,\lambda)} = 0$. Note that this value is non-zero at the ``wavefunction'' level, but the contributions from the two branches cancel once we take $|\Psi|^2$ to compute the probability distribution.

\subsection{The squeezed bispectrum with generic mass}
\label{subsec:qsfi-squeezed-completion}
We now analyze the $F_{\rm d}$ contribution for generic mass at ${\cal O}(g^2)$. For a generic $\sigma$ mass the time integrals remain Hankel integrals, but summing the branches still gives an important simplification: \eqref{eq:probability-diagonal-branch-sum} shows that only $\Gamma_{+++}$ and $\Gamma_{++-}$ are independent. In the squeezed limit $p=k_L\ll k_S=k$, the nonanalytic part of the complete $F_{\rm d}$ kernel has the universal form 
\begin{align}
 \left.\Gamma^{(3,g^2;{\rm d})}_{\tilde\rho}(p,k,k)\right|_{\rm non-analytic}
  &=\frac{c_{\rm d}g^2k}{H^2}
 \left\{C_-(\nu)x^{1/2-\nu}[1+O(x^2)]
       +C_+(\nu)x^{1/2+\nu}[1+O(x^2)]\right\}\ ,
 \nonumber\\
 &\mbox{where }\ \  x= \frac{p}{k}\ ,\;\    \nu=\sqrt{\frac94-\frac{m^2}{H^2}}\ .
 \label{eq:probability-generic-nonanalytic-structure}
\end{align}

The mass dependent coefficients are given in Appendix~\ref{app:nonanalytic-complete}. After removing the prefactor $k^2/p$, these are the usual de Sitter weights $(p/k)^{3/2\pm\nu}$.  They are useful because a local derivative expansion produces only the analytic part of the soft momentum expansion, but it is the non-integer powers, or the logarithmic oscillations for $\nu=i\mu$, that retain information about the propagating massive field. At the conformal value $\nu=1/2$ the powers become integers and this distinction disappears, which is why \eqref{eq:probability-conformal-g2} reduces to a rational function of the momenta.

The squeezed limit is studied by first taking the $\tau_0 \to 0$ limit at fixed nonzero external momenta. We then require corrections from small $p$ expansion on the soft leg to be small. At masses of order $H$, a sufficient hierarchy is $|g|\ll p\ll k$. Beyond this, a formal expansion of the order-$g^2$ coefficient for a fixed $g$ at $p\to0$ is no longer accurate. 

The trace $\int { \cal D} \bar{\sigma} $ leaves a finite contribution to the light field correlations even though $\sigma$ decays at late times. To see this, consider the part of the squeezed configuration in which the soft mode couples to the massive field at time $\tau'$, before the massive field interacts with the two hard modes at time $\tau>\tau'$. For generic real $0<\nu<3/2$, massive field propagation involves two late-time solutions, $(-\tau)^{3/2-\nu}$ and $(-\tau)^{3/2+\nu}$, which decay at different rates.

The difference between $G^\sigma_{++,p}(\tau,\tau')$ and Dirichlet propagators shows how these two solutions enter the calculation. First impose the Dirichlet condition at finite $\tau_0$, and then take $\tau_0\to0$ at fixed interaction times. In the ordering $\tau>\tau'$, suppressing overall normalizations and common powers of time, we obtain
\begin{align}
 G^\sigma_{++,p}(\tau,\tau')
 &\propto H_\nu^{(1)}(-p\tau)H_\nu^{(2)}(-p\tau')\ ,\\
 G^\sigma_{D,p}(\tau,\tau')
 &\propto J_\nu(-p\tau)H_\nu^{(2)}(-p\tau')\ .
 \label{eq:squeezed-propagator-falloffs}
\end{align}
For masses of order $H$, the hard interaction occurs at $|\tau|\sim1/k$, so $z=-p\tau\sim p/k\ll1$. The leading powers are
\begin{align}
 H_\nu^{(1)}(z)&\sim c_-z^{-\nu}+c_+z^\nu\ ,
 \qquad
 J_\nu(z)\sim c_Jz^\nu\ .
 \label{eq:squeezed-small-argument}
\end{align}
Restoring the common factor $(-\tau)^{3/2}$, we see that $G^\sigma_{++,p}(\tau,\tau')$ contains both decay laws in its later time argument, whereas the Dirichlet subtraction cancels the more slowly decaying contribution. Thus the Dirichlet bulk term alone does not give the complete squeezed signal.

Integrating over the final heavy-field value restores the subtracted contribution and generates the propagators connecting the two contour branches. Appendix~\ref{app:nonanalytic-complete} evaluates the resulting coefficients and shows that the remaining time integral contains only integer powers of $p/k$. Therefore they cannot replace the missing fractional power term. Although we organize our calculation by separating the solution into a Dirichlet contribution and a boundary contribution, the observable depends on their sum.

The two massive field solutions give the characteristic factors $x^{3/2\pm\nu}$, where $x=p/k$. For our source, an additional factor $k^2/p$ changes these to the kernel's $x^{1/2\pm\nu}$ dependence. This source factor has a soft pole and should not be called analytic at $p=0$. For $\nu=i\mu$, analytic continuation of the complete kernel gives oscillations in $\ln x$. These mass-dependent fractional powers and oscillations cannot be reproduced by any finite local derivative expansion.

\subsection{The heavy-mass limit and the bispectrum}
\label{subsec:qsfi-heavy-mass}

The terms \eqref{eq:qsfi-boundary-gaussian-result} generated by the trace over $\bar\sigma$ redshift away in the local heavy-mass expansion for our sources, assuming $m\gg H$ and adiabatic evolution\cite{Cespedes2012hu,Achucarro2012sm,Achucarro2012Decoupling}. We show this at tree level, at orders $g^2$ and $\lambda g^3$, considering expansions in the inverse powers of $m$.

For $k/a_0\ll m$, together with adiabatic assumption, the heavy mode has mode function 
\begin{equation}
    u_k(t)\simeq \frac{e^{\,i\int^t\omega_k dt'}}{\sqrt{2a^3(t)\omega_k(t)}} \ , \qquad \omega_k = \sqrt{m^2 + \frac{k^2}{a^2(t)}} \ .
\end{equation}
The adiabatic assumption is precisely $|\dot\omega| \ll \omega^2$. This leads to 
\begin{align}
    P_k \simeq\frac{1}{2a_0^3\omega_k} \ , \qquad
    I_{\pm,\mathbf k} &\simeq\mp\frac{ia_0}{\omega_k}F_{\pm,\mathbf k}(\tau_0) \ .
\end{align}
Substitute into \eqref{eq:qsfi-boundary-saddle} and \eqref{eq:qsfi-boundary-gaussian-result}, taking leading order in heavy mass expansion yields
\begin{align}
    \bar\sigma_{*,\mathbf k} &\simeq\frac{g}{2a_0^2\omega_k^2} (F_++F_-)_{\mathbf k}(\tau_0) \ , \nonumber\\
    \Delta\ln\rho_{\partial} &\simeq\frac{g^2}{4a_0}\int_{\mathbf k} \frac{(F_++F_-)_{\mathbf k}(F_++F_-)_{-\mathbf k}} {\omega_k^3}\bigg|_{\tau_0} \ .
    \label{eq:heavy-boundary-saddle-estimate}
\end{align}
Up to field-independent normalization, this is the difference between tracing over $\bar\sigma$ and fixing $\bar\sigma=0$. At late time, $\zeta'=\mathcal{O}(a_0^{-1})$. The cubic part of $\Delta \ln \rho_\partial$ therefore falls as $a_0^{-3} \propto \tau_0^3$ for $F_{\rm d}$ and as $a_0^{-2} \propto \tau_0^2$ for $F_{\rm K}$, as $-(\nabla\zeta)^2=\mathcal{O}(1)$. Further corrections in powers of $H/m$ carry the same late-time suppression. One can further check that all terms involved in $\Delta \rho_{\partial}$ have positive power of $\tau_0$. At fixed momenta and derivative order, their integrals converge and vanish as $\tau_0\to0$. 

To check the $\lambda g^3$ term, write 
\begin{equation}
    \sigma^{(0)}_{+,\mathbf{k}}(\tau) = (\bar\sigma_{\mathbf{k}} -igP_k I_{+,\mathbf{k}}) K_{+,k}(\tau) + g\int_{{\cal C}_+} d\tau' \, a^2(\tau')G_{F,k}(\tau,\tau')F_{+,\mathbf{k}}(\tau') \ .
\end{equation}
Define the homogeneous part $\sigma_{h,\mathbf{k}}(\tau)\equiv(\bar\sigma_{\mathbf{k}} -igP_k I_{+,\mathbf{k}}) K_{+,k}(\tau)$, and the particular part $\sigma_{p,\mathbf{k}}(\tau) = \sigma_{\mathbf{k}}(\tau) - \sigma_{h,\mathbf{k}}(\tau)$. Using similar approximations, we obtain that in the integral of $\int d\tau \ a^4 \sigma^3$, only the pure particular part $\int d\tau \ a^4 \sigma_p^3$ survives, and the result is proportional to $1/k_t^6$. The other parts are suppressed by powers of order $\mathcal{O}(a_0^{-6})$. Heuristically, this can be understood as the homogeneous part oscillating much faster than the particular part due to a heavy mass in the phase. These estimations then justify omitting the correction coming from a specific $\bar\sigma_*$, in agreement with the late-time matching discussed in \cite{GreenSun2025}.

We may now use the in-out bulk effective action for our calculation. With ${\cal F}=a^{-2}F$, the heavy-field equation and its adiabatic particular
solution are
\begin{align}
    (m^2-\Box)\sigma &=g{\cal F}-\frac{\lambda}{2}\sigma^2\ , & \sigma_{\rm loc}^{(0)} &=\frac{g}{m^2}{\cal F} +\frac{g}{m^4}\Box{\cal F}+\cdots \ .
    \label{eq:heavy-particular-expansion}
\end{align}
Substituting into the action gives
\begin{align}
    \Delta S_{\rm loc} =\int d^4x\sqrt{-g}\left[ \frac{g^2}{2m^2}{\cal F}^2 +\frac{g^2}{2m^4}{\cal F}\Box{\cal F} -\frac{\lambda g^3}{6m^6}{\cal F}^3+\cdots\right]\ .
    \label{eq:heavy-local-action}
\end{align}
For our two choices of $\cal F$, the leading cubic terms are
\begin{align}
    \left.\Delta S_{\rm loc}^{(g^2;{\rm d})}\right|_{\zeta^3} &=\frac{3  c_{\rm d} g^2}{m^2}\int d\tau\,d^3x\,\zeta\zeta'^2 \ , \nonumber\\
    \left.\Delta S_{\rm loc}^{(g^2;{\rm K})}\right|_{\zeta^3} &=\frac{  c_{\rm K} g^2}{m^2}\int d\tau\,d^3x\, [\zeta'^3-\zeta'(\nabla\zeta)^2] \ , \nonumber\\
    \left.\Delta S_{\rm loc}^{(\lambda)}\right|_{\zeta^3} &=-\frac{\lambda g^3}{6m^6} \int d\tau\,d^3x\,a^{-2}\zeta'^3 \ .
    \label{eq:heavy-local-cubic-actions}
\end{align}
For $K_k^\zeta=(1-ik\tau)e^{ik\tau}$,
\begin{align}
    \operatorname{Im}\int_{{\cal C}_+}d\tau\, K_{k_i}^\zeta K_{k_j}^{\zeta\prime}K_{k_\ell}^{\zeta\prime} =k_j^2k_\ell^2\left(\frac{2}{k_t^3}+\frac{6k_i}{k_t^4}\right)\ ,
    \label{eq:heavy-Fd-contact-integral}
\end{align}
so choosing $F_{\rm d}$ leads to
\begin{align}
    \Gamma^{(3,g^2;{\rm d})}_{\tilde\rho,\,\rm loc} =\frac{24 c_{\rm d} g^2}{m^2} \sum_{{\rm cyc}(i,j,\ell)}k_j^2k_\ell^2 \left(\frac{1}{k_t^3}+\frac{3k_i}{k_t^4}\right) +{\cal O}\!\left(\frac{g^2H^2}{m^4}\right)\ .
    \label{eq:heavy-local-Fd-kernel}
\end{align}
The bispectrum is then $-\Gamma^{(3,g^2;{\rm d})}_{\tilde\rho,\,\rm loc}\prod_iP_\zeta(k_i)$. This agrees with the calculation from the full theory. 

The integrals of the $\zeta'^3$ terms in $F_{\rm K}$ and $\lambda g^3$ contributions are real:
\begin{align}
    \int_{{\cal C}_+}d\tau\,\prod_iK_{k_i}^{\zeta\prime} &=-\frac{6(k_1k_2k_3)^2}{k_t^4}\ , & \int_{{\cal C}_+}d\tau\,a^{-2}\prod_iK_{k_i}^{\zeta\prime} &=\frac{120H^2(k_1k_2k_3)^2}{k_t^6} \ .
    \label{eq:heavy-real-contact-integrals}
\end{align}
The gradient term in $F_{\rm K}$ also gives a real integral. These terms change only the phase of the wavefunctional, so their contribution to the bispectra vanish. Note that this means for $F_K$, there is no leading order contribution to the bispectrum. One can actually see that the integral $\int^{0}_{-\infty} d\tau \ \tau^n e^{ik\tau}$ under BD vacuum is real for odd $n$ and imaginary for even $n$. Therefore, changing the power of $\tau$ in the integrand without introducing proper powers of $i$ at the same time, will lead to a jump between real and imaginary results. Using this property, we can see that the arguments above remains true at every order in the $\Box/m^2$ expansion, as $\Box$ introduces extra powers of $i\tau$ only. Note that here our bilinear coupling coming from $\sigma \cal{F}$ takes the form $a^2 g \sigma \zeta'$. The mixing $a^3g\sigma\zeta'$ used in standard quasi-single-field models has a different time dependence and can instead yield a nonzero local $\dot\zeta^3$ bispectrum \cite{Chen2012LargeMass, GongPiSasaki2013Equilateral}. This is due to $a$ introducing $\tau^{-1}$ without an accompanying $i$. 

Our result also agrees with tree level calculations under heavy-mass expansion starting from the full theory. To see this, use the usual in-in formalism and expand out the $\sigma$ propagators under heavy-mass limit. One then recovers the result above at leading order.

The full result, however, contains contributions outside the heavy mass expansion series. For example, Appendix~\ref{app:nonanalytic-complete} gives the $F_{\rm d}$ squeezed terms $(p/k)^{1/2\pm i\mu}$, where $\mu=\sqrt{m^2/H^2-9/4}$. Their coefficients are exponentially suppressed at large $\mu$, but they survive at late times. The squeezed expansion must satisfy $\mu p/k\ll1$ in this limit. These terms cannot be captured by any finite order in $H/m$ \cite{GreenSun2025, DuasoPueyo2026Asymptotic}. Therefore, $\bar\sigma_*$ is still needed for the exact bispectrum.

\section{A bottom-up open EFT organization for the reduced density matrix}
\label{sec:bottom-up-open-eft}

The previous section derived the reduced density matrix by explicitly integrating out $\sigma$. We now turn to the bottom-up construction of the effective in-in theory. At a finite observation time $\tau_0$, write the reduced density matrix as
\begin{align}
    \rho_{\rm red}[\bar\zeta_+,\bar\zeta_-;\tau_0] ={}&\int_{\zeta_\pm(\tau_0)=\bar\zeta_\pm}
    {\cal D}\zeta_+{\cal D}\zeta_-\, \exp\left\{iS_\zeta[\zeta_+]-iS_\zeta[\zeta_-]+iS_{\rm IF}[\zeta_+,\zeta_-]\right\}\ ,
    \label{eq:bottomup-rho}
\end{align}
The initial state is the Bunch--Davies vacuum used above. We write $S_{\rm IF}$ for the influence action and $e^{iS_{\rm IF}}$ for the influence functional.

\subsection{Universal Schwinger--Keldysh constraints}\label{sec:SK_constraints}

The effective in-in action obtained from a unitary UV theory obeys the following three non-perturbative conditions:
\begin{align}
    S_{\rm IF}[\zeta,\zeta]={}&0 \ ,
    \label{eq:bottomup-normalization}\\
    S_{\rm IF}[\zeta_+,\zeta_-]^*={}&-S_{\rm IF}[\zeta_-,\zeta_+] \ ,
    \label{eq:bottomup-hermiticity}\\
    \operatorname{Im}S_{\rm IF}[\zeta_+,\zeta_-] \geq {}& 0 \ .
    \label{eq:bottomup-positivity}
\end{align}
They follow from the normalization, Hermiticity, and positivity of the underlying full theory. These conditions are important for Schwinger--Keldysh theories such as effective actions for dissipative fluids and time-translation breaking in open systems \cite{CrossleyGloriosoLiu2017,Hongo2019Time}. Their application to open effective theories of inflation is discussed in Refs.~\cite{AguiSalcedo2024Open,Burgess:2024eng,ColasQinTong2026Collider,Colas2025Lectures,Pajer2026Lectures}.

It is useful to introduce the Schwinger--Keldysh basis
\begin{align}
    \zeta_r = \frac{\zeta_++\zeta_-}{2}\ &,
    \qquad
    \zeta_a = \zeta_+-\zeta_-\ . 
\end{align}
Equation \eqref{eq:bottomup-normalization} implies that every term in $S_{\rm IF}$ contains at least one $a$-type variable. In particular, there can be no $\zeta_r\zeta_r$ term in a quadratic action. Causality requires the kernel multiplying $\zeta_a\zeta_r$ to be retarded. Operators and kernels must also respect the symmetries of the full dynamics and state, including the nonlinear clock symmetry when $\zeta$ is the curvature perturbation \cite{AguiSalcedo2024Open}. We use $F_r=(F_++F_-)/2$ and $F_a=F_+-F_-$ for the composite sources, and in general $F_r\neq F[\zeta_r]$.

We focus on the case where the environment $\sigma$ is linearly coupled to $F$, as in the previous section. The influence functional then depends on $F_\pm$, $S_{IF}[\zeta_+,\zeta_-] = S_{IF}[F_+,F_-]$, and obeys \eqref{eq:bottomup-normalization}--\eqref{eq:bottomup-hermiticity} with $\zeta$ replaced by $F$. The most general quadratic influence action satisfying those constraints is
\begin{align}
    S_{\rm IF}^{(2)}=\int_{x,y}F_a(x){\cal D}_R(x,y)F_r(y)+\frac{i}{2}\int_{x,y}F_a(x){\cal N}(x,y)F_a(y)\ .
    \label{eq:bottomup-general-quadratic}
\end{align}
Here
\begin{align}           
    \int_{x,y}\equiv\int^{\tau_0}d\tau_x\,d\tau_y\int d^3x\,d^3y\, a^2(\tau_x)a^2(\tau_y) \ .
    \label{eq:bottomup-xy-measure}
\end{align}
For real fields, the response and noise kernels obey
\begin{align}
    {\cal D}_R(x,y)^*={}&{\cal D}_R(x,y)\ ,
    \qquad
    {\cal D}_R(x,y)=0\quad\text{for}\quad \tau_x<\tau_y\ ,
    \label{eq:bottomup-retarded-conditions}\\
    {\cal N}(x,y)^*={}&{\cal N}(x,y)={\cal N}(y,x)\ ,\nonumber\\
    \int_{x,y}f(x){\cal N}(x,y)f(y)&\geq0
    \qquad
    \text{for every real test function }f\ .
\label{eq:bottomup-noise-conditions}
\end{align}
Thus ${\cal D}_R$ contains dispersive and dissipative response, whereas ${\cal N}$ contains the state-dependent fluctuations or noise \cite{Boyanovsky2015Effective,Boyanovsky2016Stochastic}. These kernels generally retain memory of the environment. In cosmological master equations, nonlocal contributions can control secular evolution and must be treated carefully~\cite{Brahma:2024yor}. 

Since our environment is de Sitter invariant, these kernels are more restricted than the general Schwinger--Keldysh conditions. The top-down matching identifies them with heavy-field two-point functions, so for every de Sitter Killing vector $\xi^\mu$ we have
\begin{align}
    \left({\cal L}_{\xi_x}+{\cal L}_{\xi_y}\right){\cal D}_R(x,y)=0 \ ,
    \qquad
    \left({\cal L}_{\xi_x}+{\cal L}_{\xi_y}\right){\cal N}(x,y)=0 \ .
    \label{eq:bottomup-kernel-dS-Ward}
\end{align}
Focusing on a scalar field environment, the Lie derivatives reduce to
\begin{align}
    \left[\xi^\mu(x)\partial_{x^\mu}+\xi^\mu(y)\partial_{y^\mu}\right]{\cal D}_R(x,y)=0 \ ,
    \label{eq:bottomup-kernel-scalar-Ward}
\end{align}
similarly for $\cal N$. Thus the noise kernel is an invariant bi-scalar, while the response kernel has the same invariant dependence together with its causal support.

For a generic kernel $\cal K$, translations and rotations reduce it to ${\cal K}(\tau,\tau';r)$, with $r=|\bm x-\bm y|$, and the dilation identity becomes
\begin{align}
    \left(\tau\partial_\tau+\tau'\partial_{\tau'}+r\partial_r\right){\cal K}(\tau,\tau';r)=0 \ .
    \label{eq:bottomup-kernel-dilation-position}
\end{align}
With the Fourier convention used here this is equivalently
\begin{align}
    \left( \tau \partial_\tau + \tau' \partial_{\tau'}-k\partial_k-3\right){\cal K}_k(\tau,\tau')=0 \ .
    \label{eq:bottomup-kernel-dilation-momentum}
\end{align}
Special conformal transformations impose additional differential constraints through \eqref{eq:bottomup-kernel-dS-Ward}. At finite $\tau_0$, these variations also change the observation surface, and the kernel identities must be combined with that change. The future boundary is preserved in the limit $\tau_0 \to 0$ because the linear variation is proportional to $\tau_0$ \cite{Cespedes2020Time}. Section~4.4 gives the corresponding boundary Ward identities.

At cubic order in $F$, we have 
\begin{align}
    S^{(3)}_{\text {IF}} =& \frac12 \int_{x,y,z} F_a(x)F_r(y)F_r(z)\Gamma_{arr}(x,y,z) \nonumber \\
    +& \frac i2 \int_{x,y,z} F_a(x)F_a(y)F_r(z)\Gamma_{aar}(x,y,z) \nonumber \\
    +& \frac {1}{3!} \int_{x,y,z} F_a(x)F_a(y)F_a(z)\Gamma_{aaa}(x,y,z) \ .
\end{align}
Hermiticity of the reduced density matrix requires the explicit $i$ in the $aar$ term. Since we focus on the bispectrum, we retain the quadratic and cubic terms in the systematic expansion
\begin{align}
    S_{\rm IF}=S_{\rm IF}^{(2)}+S_{\rm IF}^{(3)}+\cdots .
    \label{eq:bottomup-systematic-expansion}
\end{align}
Accordingly, the bottom-up data relevant through the bispectrum are
\begin{align}
    \left\{{\cal D}_R,{\cal N};\Gamma_{arr},\Gamma_{aar},\Gamma_{aaa}\right\} \ ,
    \label{eq:bottomup-kernel-data}
\end{align}
subject jointly to the Schwinger--Keldysh constraints, causality, positivity, and the spacetime symmetries.

\subsection{Exact matching to the quasi-single-field model}

In our conventions, $G_R=G_{++}-G_{+-}$, $G_A(x,y)=G_R(y,x)$, and $G_H=(G_{+-}+G_{-+})/(2i)$. These are respectively the retarded response, its advanced counterpart, and the symmetrized heavy-field covariance \cite{Chen2017ryl,Boyanovsky2016Stochastic}. The two bottom-up kernels in \eqref{eq:bottomup-general-quadratic} are fixed by
\begin{align}
    {\cal D}_R(x,y) =g^2G_{R,\sigma}(x,y) \ ,
    \qquad
    {\cal N}(x,y)=g^2G_{H,\sigma}(x,y) \ .
    \label{eq:bottomup-QSFI-kernel-matching}
\end{align}
Therefore,
\begin{align}
    iS_{\rm IF}^{(2)} = ig^2F_aG_{R,\sigma}F_r-\frac{g^2}{2}F_aG_{H,\sigma}F_a \ .
    \label{eq:bottomup-QSFI-ra-exponent}
\end{align}
Next, at cubic order, we have 
\begin{align}
    \Gamma_{arr}(x;y,z) =& -\lambda g^3 \int_u a_u^2\,G_A(u,x)\,G_R(u,y)\,G_R(u,z) \ , \\
    \Gamma_{aar}(x,y;z) =& -\lambda g^3\int_u a_u^2\,\Big[G_A(u,x)G_H(u,y) +G_A(u,y)G_H(u,x)\Big]G_R(u,z) \ , \\
    \Gamma_{aaa}(x,y,z) =& \ \lambda g^3\int_u a_u^2\Big[G_A(u,x)G_H(u,y)G_H(u,z) +G_A(u,y)G_H(u,x)G_H(u,z)\nonumber\\
    &\qquad \qquad \ +G_A(u,z)G_H(u,x)G_H(u,y) -\frac{1}{4}G_A(u,x)G_A(u,y)G_A(u,z)\Big]\ .
\end{align}
Both the quadratic and cubic terms in $F$ therefore contribute to the bispectrum of $\zeta$.

The effective Schwinger--Keldysh action admits three different decompositions that relate to different questions. First, in the top-down calculation we can split between the $\bar\sigma$-free action and the result of tracing over the common final heavy-field value,
\begin{align}
    S_{\rm eff}^{\rm SK}
    =S_{\bar\sigma = 0} + S_{\bar\sigma\text{-trace}} \ .
    \label{eq:bottomup-fixed-trace-split}
\end{align}
At quadratic order the second term is precisely the gaussian contribution in \eqref{eq:qsfi-boundary-gaussian-result},
\begin{align}
    iS_{\bar\sigma\text{-trace}}^{(2)}
    =-\frac{g^2}{2}\int_{\mathbf k}P_k\,
    \Delta I_{\mathbf k}\Delta I_{-\mathbf k} \ .
    \label{eq:bottomup-trace-quadratic}
\end{align}
This is a natural way to choose the bulk-boundary decomposition, with the term~\eqref{eq:bottomup-trace-quadratic} being the ``boundary'' correction. However, keep in mind that there are underlying bulk propagations in $\Delta I$. 

Second, the same action can be reorganized into an in-out-like same-branch part and a non-factorizable influence action part. Equation \eqref{eq:qsfi-Feynman-kernel} shows that a diagonal piece of \eqref{eq:bottomup-trace-quadratic} completes the Dirichlet propagator to the Feynman propagator, while the remaining piece produces the mixed-branch term in \eqref{eq:qsfi-gaussian-SK-form}. Thus part of what is operationally generated by the final trace is absorbed into the ordinary same-branch factorizable description.

Third, in the $r/a$ basis we can separate the unitary direction from dissipative and noise directions. At quadratic order the time-symmetric part of ${\cal D}_R$ belongs to the unitary part, whereas its antisymmetric part and the ${\cal N}$ term describe dissipative and stochastic effects. At cubic order,
\begin{align}
    \frac13(F_+^3-F_-^3)=F_aF_rF_r+\frac{1}{12}F_aF_aF_a
    \label{eq:bottomup-cubic-unitary-direction}
\end{align}
fixes the unitary direction of the local, non-derivative $arr$ and $aaa$ vertices. Orthogonal combinations encode open-system data \cite{AguiSalcedo2024Open}. 

\subsection{Symmetry constraints on reduced density matrices}
Symmetry guides the construction of bottom-up effective theories. The same idea applies to reduced density matrices. We first discuss the general conditions under which a symmetry survives the trace and then apply them to de Sitter isometries.

Consider the reduced density matrix $\rho_{\rm red}$. Suppose the full theory has an exact symmetry $g$. The simplest case is that $g$ transforms $\zeta$ and $\sigma$ separately, that is 
\begin{align}
   \zeta \to \zeta + \delta \zeta[\zeta ]\ , \quad \sigma \to \sigma + \delta \sigma[\sigma] \ .
\end{align}
If the full state is invariant, the transformation of $\sigma$ can then be absorbed into the heavy-field trace, leaving the reduced state invariant under the transformation of $\zeta$. The trace must then be mapped into itself under the transformation. The full action, path-integral measure, regulator, state, and contour prescription must also respect the symmetry. Under these assumptions, applying the symmetry transformations to the heavy integration variables gives
\begin{align}
    \rho_{\rm red}^{\,g\Sigma} [\bar\zeta_+^{\,g},\bar\zeta_-^{\,g}] = \rho_{\rm red}^{\,\Sigma} [\bar\zeta_+,\bar\zeta_-] \ ,
    \label{eq:hypersurface-covariance}
\end{align}
where we use $\Sigma$ to label the observation surface. This result holds as long as the full state is invariant. For a spacetime symmetry, both the fields and the observation surface must be transformed. If $g\Sigma=\Sigma$, \eqref{eq:hypersurface-covariance} reduces to invariance on that surface. Note that, although the  microscopic bulk action on two branches can admit independently chosen symmetry parameters, the final trace constrains transformations acting on both branches to be the same. The minus sign in the SK action does not reverse the symmetry transformation on the $-$ branch. After the heavy field is integrated out, branch mixing terms should satisfy the diagonal transformation, but are in general no longer invariant under other combinations of $+$ and $-$ transformations. This symmetry-breaking pattern is discussed in \cite{Hongo2019Time}.

The same point can also be seen for the influence action,
\begin{align}
    S_{\rm IF}^{\,g\Sigma} [\zeta_+^g,\zeta_-^g] =S_{\rm IF}^{\,\Sigma}[\zeta_+,\zeta_-]\ ,
    \label{eq:influence-functional-symmetry}
\end{align}
up to possible boundary phases discussed below. Since this is a constraint for the full influence action, an approximation to $S_{\rm IF}$ must therefore retain all terms related by it at the order being considered.

Suppose the symmetry changes a fixed-boundary wavefunctional by a field-dependent phase,
\begin{align}
    \Psi_{g\Sigma}[\bar\zeta^g,\bar\sigma^g] =e^{\mathrm{i} B_g[\bar\zeta,\bar\sigma]} \Psi_\Sigma[\bar\zeta, \bar\sigma] \ .
    \label{eq:wavefunction-boundary-phase}
\end{align}
After tracing out $\bar\sigma$,
\begin{align}
    \rho_{\rm red}^{\,g\Sigma}[\bar\zeta_+^g,\bar\zeta_-^g] ={}& \int{\cal D}\bar\sigma\, e^{\mathrm{i} B_g[\bar\zeta_+,\bar\sigma] -\mathrm{i} B_g[\bar\zeta_-,\bar\sigma]} \Psi[\bar\zeta_+,\bar\sigma] \Psi^*[\bar\zeta_-,\bar\sigma]\ .
    \label{eq:reduced-boundary-phase}
\end{align}
A phase depending only on $\bar\sigma$ cancels between the two branches. If $B_g=B_{g,\zeta}+B_{g,\sigma}$, only the light-field phase remains,
\begin{align}
    \rho_{\rm red}^{\,g\Sigma}[\bar\zeta_+^g,\bar\zeta_-^g] ={}&e^{\ii B_{g,\zeta}[\bar\zeta_+] -\ii B_{g,\zeta}[\bar\zeta_-]} \rho_{\rm red}^{\,\Sigma}[\bar\zeta_+,\bar\zeta_-]\ .
    \label{eq:factorized-boundary-phase}
\end{align}
This is the usual unitary transformation of a density matrix. A genuinely mixed phase $B_g[\bar\zeta,\bar\sigma]$ remains inside the heavy-field integral and generally does not reduce to a light-field-only phase. For a boundary term depending only on the field configurations, however, the phases \eqref{eq:reduced-boundary-phase} cancel point by point on the diagonal $\bar\zeta_+=\bar\zeta_-$. Boundary terms involving momenta or normal derivatives instead change the boundary variables and must be treated as canonical transformations.

Thus exact integration over $\sigma$ preserves the symmetry whenever the transformation closes on $\zeta$, the full dynamics and state are invariant, the heavy-field trace is mapped into itself, and the observation surface is transformed consistently. We now apply these statements to de Sitter, where dilations and special conformal transformations move a finite-time slice but preserve the future boundary.

\subsection{De Sitter isometries and future-boundary Ward identities}
In the flat patch of de Sitter space,
\begin{align}
    \rd s^2 = \frac{1}{H^2\tau^2} \left(-\rd\tau^2+\rd\bm{x}^2\right)\ ,
    \label{eq:dS-metric}
\end{align}
an infinitesimal isometry is generated by a Killing vector $\xi^\mu$, which specifies the corresponding coordinate displacement. Suppose a bulk scalar transforms as $\delta_\xi\zeta(x)=-\xi^\mu(x) \partial_\mu \zeta(x)$. We use the boundary fields and density-matrix kernels already defined in Section~\ref{subsec:qsfi-probability-cubic-kernel}. The symmetry statements below assume an invariant action and state. The clock background must also be transformed when it is not preserved by the isometry.

Translations and rotations preserve every constant-$\tau$ surface. Dilations and special conformal transformations are generated by
\begin{align}
    \xi_D ={}&\lambda \left(\tau\partial_\tau+x^i\partial_i\right) \ , \nonumber\\
    \xi_{K_{\bm b}} ={}&2(\bm b\cdot\bm x)\tau\partial_\tau +\left[ 2(\bm b\cdot\bm x)x^i +(\tau^2-\bm x^2)b^i \right]\partial_i \ .
    \label{eq:dS-dilation-SCT-vectors}
\end{align}
At finite $\tau_0$, a dilation changes the observation time and a special conformal transformation maps a constant-time surface to a position-dependent surface. Their exact finite-time content is therefore the hypersurface covariance relation \eqref{eq:hypersurface-covariance}. The observation surface must be transformed together with the fields. Holding $\tau_0$ fixed would omit this part of the Ward identity \cite{Cespedes2020Time}.

The future boundary $\tau=0$ is preserved under all dS isometries. Suppose a scalar mode behaves as
\begin{align}
    \zeta(\tau,\bm x) \sim(-\tau)^{\Delta_\zeta} \bar\zeta(\bm x) \ .
    \label{eq:future-boundary-mode}
\end{align}
Here $\Delta_\zeta$ is the late-time scaling exponent, which is zero for the growing mode of a massless scalar. The boundary coefficient transforms as
\begin{align}
    \delta_D\bar\zeta ={}&-\lambda \left(\Delta_\zeta+\bm x\cdot\bm\nabla\right) \bar\zeta\ ,
    \label{eq:boundary-dilation-transformation} \\
    \delta_{K_{\bm b}}\bar\zeta ={}&- \left[ 2(\bm b\cdot\bm x) \left(\Delta_\zeta+\bm x\cdot\bm\nabla\right) -\bm x^2\bm b\cdot\bm\nabla \right] \bar\zeta\ .
    \label{eq:boundary-SCT-transformation}
\end{align}
For an invariant state, without boundary anomaly, the future-boundary reduced density matrix obeys the diagonal two-copy Ward identity
\begin{align}
    \sum_{s=\pm} \int\rd^3x\, \delta_\xi\bar\zeta_s(\bm x) \frac{\delta}{\delta\bar\zeta_s(\bm x)} \rho_{\rm red}[\bar\zeta_+,\bar\zeta_-] =0 \ .
    \label{eq:future-boundary-functional-Ward}
\end{align}
Both branches transform in the same direction. If a symmetry-generated boundary phase is present, its branch-antisymmetric insertion appears on the right-hand side of \eqref{eq:future-boundary-functional-Ward}. As discussed before, a configuration-only phase cancels in the diagonal probability functional.

For general order $N$, let $\Gamma_{s_1\cdots s_N}^{(N)}$ denote the coefficients of $-\ln\rho_{\rm red}$ with the same $1/N!$ normalization and momentum measure as in \eqref{eq:probability-branch-cubic-definition}. As in Section~\ref{subsec:qsfi-probability-cubic-kernel}, $\Gamma$ excludes the momentum-conserving delta function. Translation and rotation invariance require momentum conservation and rotationally invariant kernels. Conjugation exchanges the two branch labels, as in \eqref{eq:probability-hermiticity-cubic}.

For each kernel, eliminate $\bm k_N=-\sum_{a=1}^{N-1}\bm k_a$ using momentum conservation. We are then left with $N-1$ independent momenta. Dilation invariance then gives
\begin{align}
    \left[ \sum_{a=1}^{N-1} \bm k_a\cdot\frac{\partial}{\partial\bm k_a} -3+N\Delta_\zeta \right] \Gamma_{s_1\cdots s_N}^{(N)} =0 \ .
    \label{eq:kernel-dilation-identity}
\end{align}
Hence a kernel is homogeneous of degree $3-N\Delta_\zeta$.

The momentum-space special conformal identity is
\begin{align}
    \sum_{a=1}^{N-1} \left[ -2\Delta_\zeta \frac{\partial}{\partial k_a^i} +k_a^i \frac{\partial^2}{\partial k_a^j\partial k_a^j} -2k_a^j \frac{\partial^2}{\partial k_a^j\partial k_a^i} \right] \Gamma_{s_1\cdots s_N}^{(N)} =0 \ .
    \label{eq:kernel-SCT-identity}
\end{align}
The same constraint can be written symmetrically in all $N$ momenta by retaining the momentum-conserving delta function and its derivatives. The common linear spacetime transformation does not mix the branch labels. Since $\bar\zeta_+$ and $\bar\zeta_-$ are independent arguments, each branch assignment therefore obeys \eqref{eq:kernel-dilation-identity} and \eqref{eq:kernel-SCT-identity}, provided there is no boundary phase. If the symmetry generates a real local phase $B_\xi[\bar\zeta]$ depending only on the boundary field configuration, the Ward identity for $-\ln\rho_{\rm red}$ instead acquires the source $-iB_\xi[\bar\zeta_+]+iB_\xi[\bar\zeta_-]$. This source cancels on the diagonal $\bar\zeta_+=\bar\zeta_-$.

On the diagonal, the probability functional $\tilde\rho_\zeta[\bar\zeta]=\rho_{\rm red}[\bar\zeta,\bar\zeta]$ has kernels
\begin{align}
    \Gamma_{\tilde\rho}^{(N)} = \sum_{s_1,\ldots,s_N=\pm} \Gamma_{s_1\cdots s_N}^{(N)}\ .
    \label{eq:probability-kernel-branch-sum}
\end{align}
This generalizes \eqref{eq:probability-diagonal-branch-sum}. Since the same Ward operator acts on every branch assignment, the sum satisfies the same identities. These are the constraints on the probability functional used to calculate equal-time field correlators.

For the growing mode of an ordinary massless scalar, $\Delta_\zeta=0$, so that the kernels have degree three under a common rescaling of their momenta, and the first term in \eqref{eq:kernel-SCT-identity} vanishes. The remaining special conformal equations further constrain the dependence on momentum ratios. Nonlocal exchange contributions can satisfy these identities when the complete dynamics and state respect the symmetry. Explicit examples exist for Massive-exchange correlators~\cite{ArkaniHamed2018kmz,Baumann2022jpr}. The free massive field kernels $G_R$ and $G_H$ in Section~\ref{sec:SK_constraints} obey de Sitter Ward identities, but the light-field kernels $\Gamma^{(N)}_{s_1\cdots s_N}$ also depend on the transformation of the source $F$ and the clock. Their behaviors can break \eqref{eq:kernel-dilation-identity} and \eqref{eq:kernel-SCT-identity} even with \eqref{eq:bottomup-kernel-dilation-momentum} satisfied by the massive kernels. Scalar-exchange correlators with interactions that break de Sitter boosts can instead be constrained with the appropriate reduced set of symmetries \cite{Pajer:2020wxk,Jazayeri:2021fvk,Pimentel2022Boostless}.

If $\zeta$ is instead the curvature perturbation associated with an inflationary clock, dilations and special conformal transformations are nonlinearly realized~\cite{Green:2020ebl},
\begin{align}
    \delta_D\zeta ={}&-\lambda \left(1+\bm x\cdot\bm\nabla\zeta\right) \ , \nonumber\\
    \delta_{K_{\bm b}}\zeta ={}&-2\bm b\cdot\bm x -\left[ 2(\bm b\cdot\bm x)x^i-\bm x^2b^i \right]\partial_i\zeta \ .
    \label{eq:nonlinear-zeta-transformations}
\end{align}
Their field-independent shifts relate different orders rather than imposing a homogeneous identity at every order \cite{Maldacena2002vr,Creminelli2012ed,Pimentel2013gza}. In particular, the dilation identity takes the branch-resolved soft form
\begin{align}
    \sum_{r=\pm}\lim_{\bm q\to0} \Gamma_{r\,s_1\cdots s_N}^{(N+1)} (\bm q,\bm k_1,\ldots,\bm k_N) ={}& \left[ 3-\sum_{a=1}^{N-1} \bm k_a\cdot\frac{\partial}{\partial\bm k_a} \right] \Gamma_{s_1\cdots s_N}^{(N)} \ .
    \label{eq:nonlinear-dilation-soft-identity}
\end{align}
The special conformal identity fixes the first derivative of the soft coefficient,
\begin{align}
    \sum_{r=\pm}\lim_{\bm q\to0} \frac{\partial}{\partial q^i} \Gamma_{r\,s_1\cdots s_N}^{(N+1)} (\bm q,\bm k_1,\ldots,\bm k_N) ={}&\frac{1}{2}\sum_{a=1}^{N-1} \left[ k_a^i\frac{\partial^2}{\partial k_a^j\partial k_a^j} -2k_a^j \frac{\partial^2}{\partial k_a^j\partial k_a^i} \right] \Gamma_{s_1\cdots s_N}^{(N)} \ .
    \label{eq:nonlinear-SCT-soft-identity}
\end{align}
We sum over the two choices of $r$ because the physical symmetry acts diagonally on the two histories. In both soft identities, eliminate $\bm k_N=-\bm q-\sum_{a=1}^{N-1}\bm k_a$ before taking the soft limit or the derivative with respect to $\bm q$. Boundary phases and symmetry-breaking local counterterms can add explicit local inhomogeneous terms. Integrating out the heavy field preserves the nonlinear identities when the full dynamics and state respect the symmetry. A covariant action alone does not guarantee a de Sitter invariant state with a fixed clock background.

As an example, our conformal benchmark at order $g^2$ gives the two point correction
\begin{equation}
    \Gamma_{+-}^{(2)} = -\frac{g^2 k }{8H^2} \ .
\end{equation}
This does not satisfy the condition \eqref{eq:kernel-dilation-identity}. Together with $\Gamma_{++-}^{(3)}$ (and $\Gamma_{-+-}^{(3)}$ induced from conjugation and momentum permutation), condition \eqref{eq:nonlinear-dilation-soft-identity} is not satisfied either. One then conclude that our model here does not include all necessary terms for $\zeta$ to be curvature perturbations. Note that these conditions are to be applied when there is a well-defined boundary kernel. Our kernels have controlled boundary values due to our choice of clock which results in an $a^2 g \sigma \zeta'$ mixing. For the bilinear mixing $a^3 g \sigma \zeta'$ used in other works, a single $\Gamma_{+-}$ would contain large logs that diverges at $\tau_0 \to 0$. However, in the full entry $\Gamma^{(2)} = \Gamma_{++}^{(2)} + \Gamma_{--}^{(2)} + \Gamma_{-+}^{(2)} + \Gamma_{+-}^{(2)}$, the log divergences cancel, and $\Gamma^{(2)}$ now satisfies \eqref{eq:kernel-dilation-identity}. This cancellation is consistent with the broader observation that equal-time observables can be simpler than their wavefunction building blocks~\cite{AguiSalcedo:2023nds,Arkani-Hamed:2025mce,Chowdhury:2026dwm}.

\section{Discussion and conclusion}\label{sec:conclusion}
We have developed the prescription to systematically construct the reduced density matrix of a light scalar by tracing out a massive field at a finite observation time. We first solve for the massive field at a fixed final value and then formalize the prescription of tracing over this value. The trace generates the mixed-branch terms required by the Schwinger--Keldysh description. This extends the finite-time matching discussed in~\cite{GreenSun2025,Cespedes2025Quartic} to derivative light-field sources and a cubic massive field self-interaction. We keep the full mass dependence and non-local structures at this stage. For $F_{\rm d}$, the conformal benchmark agrees with a direct calculation in the original two-field theory at both coupling orders, consistently with the relation between wavefunctional and in-in calculations~\cite{Chen2017ryl,Palma2026Map}. 

The calculation shows why decay of the massive field at late times does not by itself justify discarding its final trace. The homogeneous solution is normalized at the observation surface, and this normalization can compensate the decay of the boundary fluctuations. The conformal benchmark shows an explicit finite contribution to the light-field probability, even though the powers of $x=p/k$ in the squeezed limit are integers for $\nu=1/2$. At order $g^2$ with generic real $0<\nu<3/2$, the trace restores the slower falloff removed from the later-time argument of the Dirichlet propagator as $\tau_0\to0$. Analytic continuation of the complete kernel gives logarithmic oscillations for $m>3H/2$, within the squeezed regime specified in Section~\ref{subsec:qsfi-squeezed-completion}. This connects the trace construction to the massive-field signatures studied in quasi-single-field inflation and cosmological collider physics~\cite{Chen2009we, Chen2009zp, ArkaniHamed2015bza, Chen2016QuantumClocks}. Both the bulk and trace-generated contributions are necessary for calculating the observable.

We also studied the heavy mass expansion, which lead to a local effective action given a scale hierarchy and adiabatic evolution~\cite{Cespedes2012hu, Achucarro2012sm, Achucarro2012Decoupling}. For our sources with $a^2$ mixing, the trace-generated boundary corrections redshift away order by order in this expansion. The leading local action gives a nonzero $F_{\rm d}$ probability kernel. The local $F_{\rm K}$ contribution at order $g^2$ and the common contribution at order $\lambda g^3$ instead change only the late-time wavefunctional phase. These cancellations within the derivative series do not exclude an exponentially suppressed nonlocal signal, consistent with the limitations of cosmological heavy-mass expansions~\cite{DuasoPueyo2026Asymptotic}. They also depend on the time dependence of the interactions. The constant cosmic-time mixing considered in Appendix~\ref{app:a3-mixing} permits nonzero local probability kernels, as in the heavy-mass limits of quasi-single-field models~\cite{Chen2012LargeMass, NoumiYamaguchiYokoyama2013, GongPiSasaki2013Equilateral}. 

The effects of the heavy degrees of freedom are encoded in the influence action for an open effective description, similar to the discussions in~\cite{Boyanovsky2016Stochastic, Boyanovsky2015Effective, Colas2022Open, AguiSalcedo2024Open, ColasQinTong2026Collider, Colas2025Lectures, Pajer2026Lectures}. A nonlinear light-field source produces a bispectrum even when the influence action is Gaussian in the source. The massive field self-interaction generates additional nonlinear response and non-Gaussian fluctuations. Keeping these kernels nonlocal allows the reduced theory to retain the mass dependence without assuming $m\gg H$. 

Lastly, we studied the bottom-up construction of the effective description, and discussed the symmetry constraints. When a symmetry of the full theory acts on the effective description while preserving the partial trace, the reduced density matrix obeys diagonal Ward identities, with both branches transformed together. At finite time, the observation surface must also be transformed~\cite{Cespedes2020Time}.

Starting from our prescription, there are plenty of interesting future directions. One further step is to apply this formalism to realistic inflationary scenarios~\cite{Cheung2007st, Maldacena2002vr, Creminelli2012ed, Pimentel2013gza}, so that the kernels can be interpreted directly in terms of curvature-perturbation observables. It would also be interesting to relate the finite time matching developed here to the effective theory of superhorizon modes and the renormalization-group evolution of their reduced density matrix studied in~\cite{Cohen:2020php,Green:2025hmo}. 

Our calculation assumes the Bunch--Davies state and is restricted to tree level at leading orders. Extending it to loops requires treating bulk and boundary fluctuations, contractions generated by the final trace, and counterterms together. The distinction between wavefunctional loops and contractions of wavefunctional coefficients is relevant here~\cite{Weinberg2005vy, Cespedes2023IR, Qin:2024gtr, Palma2026Map}. Further extending it to a specified non-Bunch--Davies initial state would help us determine how initial excitations modify the influence action and the observables, building on existing works such as~\cite{Agarwal:2012mq,Yin:2023jlv,Chopping:2024oiu,Ghosh:2025pxn}.

It would also be useful to relax the perturbative treatment of the quadratic mixing. Recent exact solutions for cosmological collider signals at strong mixing provide a starting point~\cite{Huenupi:2026abj,Huenupi:2026aqc,Wang:2026lff,Pinol:2026xnl}. Applying the finite-time trace to these solutions would allow us to study the reduced state and its non-Gaussian correlations beyond weak mixing. 

The symmetry discussion suggests other interesting discussions. It would be interesting to combine these identities with the Schwinger--Keldysh consistency conditions and the methods developed for the cosmological bootstrap and cutting relations~\cite{ArkaniHamed2018kmz, Goodhew2020hob, GoodhewJazayeriLeePajer2021, Baumann2022jpr}. One could then ask how much of the mixed-branch kernels is fixed by symmetry, singularities, and factorization, and which information must still be supplied by the state and the final trace. For interactions that break de Sitter boosts, this construction should use the surviving symmetries~\cite{Pimentel2022Boostless}. The explicit kernels obtained here provide examples on which to test these constraints. It would also be useful to determine how the reality and phase constraints associated with cosmological CPT are inherited by the reduced density matrix~\cite{Goodhew:2024eup,Thavanesan:2025kyc}.

Applying this formalism to massive spinning fields would allow us to derive influence actions with nontrivial angular and helicity dependence~\cite{Lee:2016vti,Pimentel2022Boostless}. This would provide further tests of our matching prescription. Further exploration also includes extending to light spinning fields coupled with the inflationary background and studying correlations between massless gauge and tensor fields~\cite{Bordin:2018pca,Baumann:2020dch}. These are perfect examples for studying how the corresponding spacetime symmetries and gauge constraints are realized in the reduced effective description.

\subsubsection*{Acknowledgments}
We thank Shuntaro Aoki, Dong-Gang Wang, and Zhong-zhi Xianyu for useful and interesting discussions. G.S. is supported by the National Natural Science Foundation of China under Grant No. 12547101. S.W. is supported by grant No. 9610650 from APRC-CityU New Research Initiatives/Infrastructure Support from Central and ECS Grants No. 21313926 from the Research Grants Council, University Grants Committee. We used ChatGPT by OpenAI to assist with calculations and manuscript preparation. The authors reviewed the resulting text and verified the calculations.

\clearpage
\begin{appendices}

\section{Mode functions and explicit three-point checks}
\label{app:exact-three-point}

This appendix gives the ingredients needed to evaluate the cubic kernels in Section~\ref{subsec:qsfi-probability-cubic-kernel}. We keep the same propagator notation as in the main text.

\subsection{Mode functions and Schwinger--Keldysh propagators}

For the massless light field, after taking the late-time limit $\tau_0\to0$,
\begin{align}
    K^\zeta_{+,k}(\tau)=(1-ik\tau)e^{ik\tau} \ ,
    \qquad
    K^\zeta_{-,k}(\tau)=(1+ik\tau)e^{-ik\tau} \ .
    \label{eq:app-light-modes}
\end{align}
For the heavy field we use
\begin{align}
    \nu=\sqrt{\frac94-\frac{m^2}{H^2}} \ ,
    \label{eq:app-nu}
\end{align}
and the Bunch--Davies mode
\begin{align}
    u_{\sigma,k}(\tau) = \frac{\sqrt\pi}{2}H e^{-\frac{i\pi}{2}(\nu+\frac12)} (-\tau)^{3/2}H_\nu^{(2)}(-k\tau) \ . 
    \label{eq:app-heavy-mode}
\end{align}
The late time Schwinger Keldysh propagators are equivalently
\begin{align}
    G^\sigma_{+-,k}(\tau,\tau') ={}&i\,u_{\sigma,k}(\tau)u^*_{\sigma,k}(\tau') \ ,
    \qquad
    G^\sigma_{-+,k}(\tau,\tau') =i\,u^*_{\sigma,k}(\tau)u_{\sigma,k}(\tau') \ , \nonumber\\
    G^\sigma_{++,k}(\tau,\tau') ={}&G^\sigma_{+-,k}(\tau,\tau')\theta(\tau'-\tau) +G^\sigma_{-+,k}(\tau,\tau')\theta(\tau-\tau') \ ,
    \qquad
    G^\sigma_{--,k}=(G^\sigma_{++,k})^* \ .
    \label{eq:app-SK-propagators}
\end{align}
Here and below the $i\epsilon$ prescription is inherited from ${\cal C}_\pm$.

\subsection{Kinetic benchmark}
\label{app:kinetic-kernels}

For completeness, the two independent order-$g^2$ kernels for $F_{\rm K}=\zeta'+c_{\rm K}[\zeta'^2-(\nabla\zeta)^2]$ are
\begin{align}
    \Gamma^{(3,g^2;{\rm K})}_{+++}(1,2,3) ={}&-2ic_{\rm K}g^2\sum_{{\rm cyc}(i,j,\ell)} \int_{{\cal C}_+}d\tau d\tau'\, a^2(\tau)a^2(\tau') K^{\zeta\prime}_{+,k_i}(\tau) G^\sigma_{++,k_i}(\tau,\tau') \nonumber\\[-1mm]
    &\hspace{16mm}\times \left[ K^{\zeta\prime}_{+,k_j}(\tau')K^{\zeta\prime}_{+,k_\ell}(\tau') +\mathbf k_j\!\cdot\!\mathbf k_\ell K^\zeta_{+,k_j}(\tau')K^\zeta_{+,k_\ell}(\tau') \right] \ ,
    \label{eq:app-FK-+++} \\
    \Gamma^{(3,g^2;{\rm K})}_{++-}(1,2;3) ={}&2ic_{\rm K}g^2 \int_{{\cal C}_+}d\tau\int_{{\cal C}_-}d\tau'\, a^2(\tau)a^2(\tau') \left[ K^{\zeta\prime}_{+,k_1}K^{\zeta\prime}_{+,k_2} +\mathbf k_1\!\cdot\!\mathbf k_2 K^\zeta_{+,k_1}K^\zeta_{+,k_2} \right](\tau) \nonumber\\[-1mm]
    &\hspace{16mm}\times G^\sigma_{+-,k_3}(\tau,\tau') K^{\zeta\prime}_{-,k_3}(\tau') \ .
    \label{eq:app-FK-++-}
\end{align}
The $+--$ and $---$ kernels follow from
\eqref{eq:probability-hermiticity-cubic}.  Substitution into
\eqref{eq:probability-diagonal-branch-sum} gives the complete diagonal
kernel. 

\subsection{Conformally coupled heavy field}
\label{app:conformal-check}

For $m^2=2H^2$ one has $\nu=1/2$ and
\begin{align}
    u_{\sigma,k}(\tau) =-\frac{H\tau}{\sqrt{2k}}e^{ik\tau} \ .
    \label{eq:app-conformal-mode}
\end{align}
The propagators therefore reduce to
\begin{align}
    G^\sigma_{+-,k}(\tau,\tau') ={}&\frac{iH^2\tau\tau'}{2k}e^{ik(\tau-\tau')} \ ,
    \qquad
    G^\sigma_{-+,k}(\tau,\tau') =\frac{iH^2\tau\tau'}{2k}e^{-ik(\tau-\tau')} \ , \nonumber\\
    G^\sigma_{++,k}(\tau,\tau') ={}&\frac{iH^2\tau\tau'}{2k} \left[ e^{-ik(\tau-\tau')}\theta(\tau-\tau') +e^{ik(\tau-\tau')}\theta(\tau'-\tau) \right] \ .
    \label{eq:app-conformal-propagators}
\end{align}
\subsubsection{The effective theory calculation}
For $F_{\rm d}$ the two light-field factors that appear in \eqref{eq:probability-Fd-+++} and \eqref{eq:probability-Fd-++-} are
\begin{align}
    a^2(\tau)K^{\zeta\prime}_{+,k_i}(\tau) ={}&\frac{k_i^2}{H^2\tau}e^{ik_i\tau} \ , \nonumber\\
    a^2(\tau)\partial_\tau \left[K^\zeta_{+,k_j}(\tau)K^\zeta_{+,k_\ell}(\tau)\right] ={}&\frac{e^{i(k_j+k_\ell)\tau}}{H^2\tau} \left[ k_j^2+k_\ell^2 -ik_jk_\ell(k_j+k_\ell)\tau \right] \ .
    \label{eq:app-conformal-light-factors}
\end{align}
The factors of $\tau\tau'$ in \eqref{eq:app-conformal-propagators} cancel the denominators in \eqref{eq:app-conformal-light-factors}. Splitting the Feynman propagator into its two time-ordering regions leaves only elementary integrals such as
\begin{align}
    \int_{-\infty(1-i\epsilon)}^0d\tau\,e^{ik_t\tau} =-\frac{i}{k_t} \ ,
    \qquad
    \int_{-\infty(1-i\epsilon)}^0d\tau\,\tau e^{ik_t\tau} =\frac{1}{k_t^2} \ ,
    \qquad
    \int_{-\infty(1-i\epsilon)}^0d\tau\,\tau^2 e^{ik_t\tau} =\frac{2i}{k_t^3} \ .
    \label{eq:app-elementary-integrals}
\end{align}
Writing $k_t=k_1+k_2+k_3$, $\Gamma_{+++}$ is given by
\begin{align}
    \Gamma^{(3,g^2;{\rm d})}_{+++}(1,2,3) =-\frac{3 c_{\rm d} g^2}{2H^2} \sum_{\rm cyc(i,j,\ell)} \Bigg[ &\frac{k_i(k_j^2+k_\ell^2)}{k_t^2} +\frac{2k_ik_jk_\ell(k_j+k_\ell)}{k_t^3} \nonumber\\[-1mm]
    &+\frac{k_j^2+k_\ell^2}{2k_t} +\frac{k_jk_\ell(k_j+k_\ell)}{2k_t^2} \Bigg] \ ,
    \label{eq:app-conformal-+++}
\end{align}
and the mixed kernel in the channel where the heavy line carries $k_i$ is
\begin{align}
    \Gamma^{(3,g^2;{\rm d})}_{++-}(k_j,k_\ell;k_i) =-\frac{3 c_{\rm d} g^2}{4H^2} \left[ \frac{k_j^2+k_\ell^2}{k_t} +\frac{k_jk_\ell(k_j+k_\ell)}{k_t^2} \right] \ .
    \label{eq:app-conformal-++-}
\end{align}
The other kernels are their branch conjugates. Adding all branches by \eqref{eq:probability-diagonal-branch-sum} gives the much simpler result
\begin{align}
    \Gamma^{(3,g^2;{\rm d})}_{\tilde\rho}(1,2,3) =-\frac{6 c_{\rm d} g^2}{H^2} \left[ k_t -\frac{k_1k_2+k_2k_3+k_3k_1}{k_t} -\frac{k_1k_2k_3}{k_t^2} \right] \ ,
\end{align}
which is Eq.~\eqref{eq:probability-conformal-g2}.

The $\lambda g^3$ term simplifies even further. For each external mode,
\begin{align}
    &\int_{{\cal C}_+}d\tau'\,a^2(\tau') G^\sigma_{++,k}(\tau,\tau')K^{\zeta\prime}_{+,k}(\tau')- \int_{{\cal C}_-}d\tau'\,a^2(\tau') G^\sigma_{+-,k}(\tau,\tau')K^{\zeta\prime}_{-,k}(\tau') =\frac{\tau}{2}(1-ik\tau)e^{ik\tau} \ .
\label{eq:app-conformal-linear-response}
\end{align}
Hence Eq.~\eqref{eq:probability-lambda-diagonal-kernel} becomes
\begin{align}
    \Gamma^{(3,\lambda)}_{\tilde\rho} =-\frac{\lambda g^3}{4H^4} \operatorname{Im} \int_{-\infty(1-i\epsilon)}^0 \frac{d\tau}{\tau} e^{ik_t\tau} \prod_{i=1}^3(1-ik_i\tau) \ .
    \label{eq:app-conformal-lambda-integral}
\end{align}
All terms obtained by expanding the polynomial are real after the time integral except the $1/\tau$ term. With the Bunch Davies prescription,
\begin{align}
    \operatorname{Im} \int_{-\infty(1-i\epsilon)}^0\frac{d\tau}{\tau}e^{ik_t\tau} =\frac{\pi}{2} \ ,
\end{align}
which immediately gives $\Gamma^{(3,\lambda)}_{\tilde\rho}=-\pi\lambda g^3/(8H^4)$.

\subsubsection{The full theory calculation}
\label{app:full-theory-check}

We now start from the original two-field theory with $F_{\rm d}=\zeta'+3 c_{\rm d}  \zeta\zeta'$, and compute the in-in expectation value directly. The canonical momentum is $\pi_\zeta=a^2[\zeta'+g\sigma(1+3  c_{\rm d} \zeta)]$, so the interaction-picture Hamiltonian is
\begin{align}
 H_I(\tau)=\int d^3x\left[  -ga^2\sigma(1+3c_{\rm d}\zeta)\zeta'_I  +\frac{g^2a^2}{2}\sigma^2(1+3c_{\rm d}\zeta)^2  +\frac{\lambda a^4}{3!}\sigma^3\right]\ ,  \qquad \zeta'_I=\frac{\pi_{\zeta,I}}{a^2}\ .  \label{eq:app-full-Hamiltonian}
\end{align}
The term quadratic in $g$ is required by the Legendre transform of the derivative coupling.  However at the order being considered in this work, it does not contribute to a connected tree-level bispectrum of $\zeta$. We evaluate
\begin{align}
    \langle\bar\zeta^3\rangle =\left\langle\overline{T}e^{i\int H_I} \bar\zeta^3 T e^{-i\int H_I}\right\rangle_0 \ .
    \label{eq:app-full-in-in}
\end{align}

It is convenient to start with the linear mixing vertex, $-ga^2\sigma\zeta'_I$. With $P_\zeta(k)=H^2/(2k^3)$, the resulting mixed contraction is
\begin{align}
    \mathcal M_k(\tau) &\equiv \left.\langle\bar\zeta_{\mathbf k} \sigma_{+,-\mathbf k}(\tau)\rangle'\right|_{O(g)} \ , \nonumber\\
    \frac{\mathcal M_k(\tau)}{ P_\zeta(k)} &=g \frac{ik\tau}{2}\left[ e^{-ik\tau}\int_{-\infty(1-i\epsilon)}^\tau ds\,e^{2iks} +e^{ik\tau}\int_\tau^0ds -e^{ik\tau}\int_{-\infty(1+i\epsilon)}^0ds\,e^{-2iks} \right] \nonumber\\
    &=\frac{g\tau}{2}(1-ik\tau)e^{ik\tau} \ .
    \label{eq:app-full-mixed-contraction}
\end{align}
The first two terms are the two time orderings on the $+$ branch; the last is the mixing vertex on the $-$ branch. Thus this result follows from the two-field Wick contractions, without using an influence functional. The contraction on the opposite branch is its complex conjugate.

At order $g^2$, one inserts the cubic Hamiltonian $-3gc_{\rm d}a^2\sigma\zeta\zeta'_I$. See the left diagram of Fig.~\ref{fig:bispectrum-mechanisms}. Contracting its $\sigma$ with each of the three external legs through the linear coupling, and summing the two contractions of $\zeta\zeta'$, gives 
\begin{align}
    \left.\langle\bar\zeta_{\mathbf k_1} \bar\zeta_{\mathbf k_2}\bar\zeta_{\mathbf k_3}\rangle_c' \right|_{g^2} &=-6 c_{\rm d} g\,\operatorname{Im}\sum_{{\rm cyc}(i,j,\ell)} \int_{-\infty(1-i\epsilon)}^0d\tau\,a^2 \mathcal M_{k_i}(\tau)P_\zeta(k_j)P_\zeta(k_\ell) \partial_\tau[K^\zeta_{+,k_j}K^\zeta_{+,k_\ell}] \nonumber\\
    &=-\frac{6 c_{\rm d}  g^2}{H^2}\prod_iP_\zeta(k_i)\, \operatorname{Im}\int_{-\infty(1-i\epsilon)}^0 \frac{d\tau}{\tau}\, \partial_\tau\prod_iK^\zeta_{+,k_i}(\tau)\ .
    \label{eq:app-full-g2-integral}
\end{align}
In the second line we used $\sum_{\rm cyc}K_i\partial_\tau(K_jK_\ell) =2\partial_\tau(K_1K_2K_3)$. Defining $Q=k_1k_2+k_2k_3+k_3k_1$ and $R=k_1k_2k_3$, the remaining integral is
\begin{align}
 \int_{-\infty(1-i\epsilon)}^0\frac{d\tau}{\tau} \partial_\tau\!\left[e^{ik_t\tau} (1-ik_t\tau-Q\tau^2+iR\tau^3)\right]
 =-i\left(k_t-\frac{Q}{k_t}-\frac{R}{k_t^2}\right)\ .
 \label{eq:app-full-g2-evaluation}
\end{align}

At order $\lambda g^3$, each external leg requires one mixing insertion, which is evaluated in \eqref{eq:app-full-mixed-contraction}. The two branches therefore give
\begin{align}
    \left.\langle\bar\zeta_{\mathbf k_1} \bar\zeta_{\mathbf k_2}\bar\zeta_{\mathbf k_3}\rangle_c' \right|_{\lambda g^3} &=2\lambda\,\operatorname{Im} \int_{-\infty(1-i\epsilon)}^0d\tau\,a^4 \prod_i\mathcal M_{k_i}(\tau) \nonumber\\
    &=\frac{\lambda g^3}{4H^4}\prod_iP_\zeta(k_i)\, \operatorname{Im}\int_{-\infty(1-i\epsilon)}^0 \frac{d\tau}{\tau}e^{ik_t\tau}\prod_i(1-ik_i\tau) =\frac{\pi\lambda g^3}{8H^4}\prod_iP_\zeta(k_i) \ .
    \label{eq:app-full-lambda-evaluation}
\end{align}
Only the $1/\tau$ term has a nonzero imaginary part $\pi/2$. Combining the two contributions reproduces \eqref{eq:probability-conformal-bispectrum} directly from the full theory, at tree level and at the stated orders in $g$ and $\lambda$.

\subsection{The squeezed limit of cubic kernel at generic mass and the heavy field trace}
\label{app:nonanalytic-complete}

We derive the coefficients $C_\pm(\nu)$ in \eqref{eq:probability-generic-nonanalytic-structure} and show explicitly how the trace supplies the missing massive field falloff.  The calculation concerns $F_{\rm d}=\zeta'+3c_{\rm d}\zeta\zeta'$ at order $c_{\rm d}g^2$, with the mixing $a^2g\sigma\zeta'$ specified in Section~\ref{sec:qsfi}.  We take the late-time limit after completing the trace, at fixed nonzero momenta, and then set $k_1=p$, $k_2=k_3=k$, and $x=p/k\ll1$.  The derivation initially assumes noninteger $0<\nu<3/2$, where $\nu=\sqrt{9/4-m^2/H^2}$.  Continuation to heavier masses and the exceptional indices are discussed below.

It is useful to sum the contributions of the linear mixing on the two branches first. Define
\begin{align}
    R_k(\tau)={}&\int_{{\cal C}_+}d\eta\,a^2(\eta)  G^\sigma_{++,k}(\tau,\eta)K^{\zeta\prime}_{+,k}(\eta)  -\int_{{\cal C}_-}d\eta\,a^2(\eta) G^\sigma_{+-,k}(\tau,\eta)K^{\zeta\prime}_{-,k}(\eta) \ .
    \label{eq:a4new-response}
\end{align}
This is the ${\cal O}(g)$ part of $\sigma_k$, corresponding to the bracket in \eqref{eq:probability-lambda-diagonal-kernel}. The second term and the completion of $G_{++}$ from $G_D$ both follow from the trace. Plugging \eqref{eq:probability-Fd-+++} and \eqref{eq:probability-Fd-++-} into \eqref{eq:probability-diagonal-branch-sum} gives the complete kernel
\begin{align}
    \Gamma^{(3,g^2;{\rm d})}_{\tilde\rho}(1,2,3) =6c_{\rm d}g^2\operatorname{Im} \sum_{{\rm cyc}(i,j,\ell)}\int_{{\cal C}_+}d\tau\,a^2(\tau) R_{k_i}(\tau) \partial_\tau\!\left[K^\zeta_{+,k_j}(\tau)K^\zeta_{+,k_\ell}(\tau)\right] \ .
    \label{eq:a4new-kernel}
\end{align}
We only need to evaluate $R_k(\tau)$ once, before performing the remaining integral.

For $z=-k\tau$, separate out the momentum dependence from the mode and response by 
\begin{align}
    u_{\sigma,k}(\tau)&=Hk^{-3/2}h_\nu(z) \ , &h_\nu(z)&=\frac{\sqrt\pi}{2}e^{-i\theta_\nu} z^{3/2}H^{(2)}_\nu(z) \ , \nonumber\\
    R_k(\tau)&=k^{-1}r_\nu(z) \ , &\theta_\nu&=\frac\pi4+\frac{\pi\nu}{2} \ .
    \label{eq:a4new-dimensionless}
\end{align}
Here $K^\zeta_{+,k}(\tau)=(1+iz)e^{-iz}$. Let $w = -k\eta$. Splitting the same-branch propagator at $\eta=\tau$ gives
\begin{align}
    r_\nu(z)=-i\biggl[& h_\nu^*(z)\int_z^\infty\frac{dw}{w}h_\nu(w)e^{-iw} +h_\nu(z)\int_0^z\frac{dw}{w}h_\nu^*(w)e^{-iw} -h_\nu(z)A_\nu^*\biggr] \ , \nonumber\\
    A_\nu&\equiv\int_0^\infty\frac{dw}{w}h_\nu(w)e^{-iw} \ .
    \label{eq:a4new-time-orderings}
\end{align}
The three terms describe mixing before the hard interaction, mixing after it, and mixing on the opposite branch, respectively. Integrals at infinity inherit the Bunch Davies $i\epsilon $ deformation.

Extending the first mixing integral to the future boundary $w=0$, and subtracting the added interval, separates the response into
\begin{align}
    r_\nu(z)&=r_{\nu,{\rm h}}(z)+r_{\nu,{\rm an}}(z)\ ,  \nonumber\\
    r_{\nu,{\rm h}}(z)&=-i\left[A_\nu h_\nu^*(z)-A_\nu^*h_\nu(z)\right]\ , \nonumber\\
    r_{\nu,{\rm an}}(z)&=i\int_0^z\frac{dw}{w} \left[h_\nu^*(z)h_\nu(w)-h_\nu(z)h_\nu^*(w)\right]e^{-iw}\ .
    \label{eq:a4new-response-split}
\end{align}
The homogeneous term $r_{\nu, {\rm h}}$ contains the two massive field falloffs. The remaining analytic part $r_{\nu, {\rm an}}$ combines the second term of $r_{\nu}$ in~\eqref{eq:a4new-time-orderings} with the integral for $0<w<z$. To see why $r_{\nu,{\rm an}}$ has only integer powers, use the massive field equation:
\begin{align}
    \left[\partial_z^2-\frac2z\partial_z+1+ \frac{m^2}{H^2z^2}\right]r_{\nu,{\rm an}}(z)&=-ze^{-iz}\ , \nonumber\\
    r_{\nu,{\rm an}}(z)&=-\frac{z^3}{m^2/H^2} +\frac{iz^4}{m^2/H^2+4}+O(z^5)\ .
    \label{eq:a4new-analytic-response}
\end{align}
More generally, inserting $\sum_{n\ge3}d_nz^n$ gives
\begin{equation*}
    \left[\frac{m^2}{H^2}+n(n-3)\right]d_n+d_{n-2}  =-\frac{(-i)^{n-3}}{(n-3)!}\ ,\qquad d_1=d_2=0\ .
\end{equation*}
For $m^2>0$ none of these recursion denominators vanishes. The finite interval in \eqref{eq:a4new-response-split} selects this analytic particular solution, so it cannot supply an additional noninteger family.

To evaluate the soft mixing integral, rotate $w=-it$, such that $h_\nu(-it)=-it^{3/2}K_\nu(t)/\sqrt\pi$. Here (and only here) $K_\nu$ is the modified Bessel function of the second kind. We then have
\begin{align}
    A_\nu&=-\frac{i}{\sqrt\pi}  \int_0^\infty dt\,t^{1/2}e^{-t}K_\nu(t)=-iq_\nu\ ,
    &q_\nu& = \frac{\Gamma(\frac32+\nu) \Gamma(\frac32-\nu)}{2\sqrt2}\ .
    \label{eq:a4new-mixing-integral}
\end{align}
The integral converges for $|\operatorname{Re}\nu|<3/2$. For physical positive masses $q_\nu$ is real, and $r_{\nu,{\rm h}}=-2q_\nu\operatorname{Re}h_\nu$. Writing $\Delta_\pm=3/2\pm\nu$, its small-argument expansion is
\begin{align}
    r_{\nu,{\rm h}}(z)  &=\sum_{s=\pm}b_s(\nu)z^{\Delta_s} \left[1-\frac{z^2}{4(1+s\nu)}+O(z^4)\right] \ , \nonumber\\
    b_-(\nu)&=-\frac{q_\nu\,2^\nu\Gamma(\nu)}{\sqrt\pi} \sin\!\left(\frac\pi4+\frac{\pi\nu}{2}\right) \ , \qquad b_+(\nu)=b_-(-\nu)\ .
 \label{eq:a4new-falloffs}
\end{align}
In the sum, $s\nu$ denotes $+\nu$ or $-\nu$.

For~\eqref{eq:a4new-kernel}, only the channel with the massive momentum being $p$ produces these noninteger powers. In the other two channels where the massive momentum is $k$, the soft external light mode has a Taylor expansion in $p/k$. With $t=-k\tau$, the soft channel is
\begin{align}
    \left.\Gamma^{(3,g^2;{\rm d})}_{\tilde\rho}(p,k,k)  \right|_{\sigma\ {\rm carries}\ p}  =-\frac{12c_{\rm d}g^2k}{H^2x}\operatorname{Im}  \int_0^\infty dt\,\frac{1+it}{t}e^{-2it}r_\nu(xt) \ .
    \label{eq:a4new-soft-channel}
\end{align}
At fixed mass, the hard interaction probes $t$ of order one, where $xt\ll1$.  Each homogeneous power then requires only the elementary hard-interaction integral
\begin{align}
    \int_0^\infty dt\,t^{\Delta-1}(1+it)e^{-2it}  =\frac{(\Delta+2)\Gamma(\Delta)}{2^{\Delta+1}} e^{-i\pi\Delta/2}\ , 
    \qquad \operatorname{Re}\Delta>0\ .
    \label{eq:a4new-hard-integral}
\end{align}
Its phase is fixed by the same Bunch--Davies prescription. Substitution into \eqref{eq:a4new-soft-channel} gives
\begin{align}
    \left.\Gamma^{(3,g^2;{\rm d})}_{\tilde\rho}(p,k,k)  \right|_{\rm nonanalytic}
    =\frac{c_{\rm d}g^2k}{H^2} \left\{C_-(\nu)x^{1/2-\nu}[1+O(x^2)] +C_+(\nu)x^{1/2+\nu}[1+O(x^2)]\right\}\ ,
    \label{eq:a4new-squeezed-result}
\end{align}
where $C_\pm=6b_\pm(\Delta_\pm+2)2^{-\Delta_\pm} \Gamma(\Delta_\pm)\sin(\pi\Delta_\pm/2)$.  In terms of $\nu$ alone,
\begin{align}
    C_-(\nu)={}&-\frac{3\,2^{2\nu-2}}{\sqrt\pi}  \Gamma\!\left(\frac32+\nu\right)  \Gamma\!\left(\frac32-\nu\right)^2\Gamma(\nu)  \left(\frac72-\nu\right)  \sin^2\!\left(\frac\pi4+\frac{\pi\nu}{2}\right)\ , \nonumber\\
    C_+(\nu)={}&C_-(-\nu)\ .
    \label{eq:a4new-explicit-coefficients}
\end{align}
The corrections within each family advance in even powers of $x$. They are separate from the ordinary Taylor terms omitted in \eqref{eq:a4new-squeezed-result}. The factor $1/x$ in \eqref{eq:a4new-soft-channel} explains the shift from the massive field weights $3/2\pm\nu$ to the kernel powers $1/2\pm\nu$. Equivalently, $k x^{1/2\pm\nu}=(k^2/p)x^{3/2\pm\nu}$. The prefactor $k^2/p$ has an integer soft power but is not analytic at $p=0$. The bispectrum follows by multiplying the complete kernel by $-P_\zeta(p)P_\zeta(k)^2$, as in \eqref{eq:probability-cubic-kernel-bispectrum}.

The role of the trace can now be shown directly. For real $0<\nu<3/2$, the late-time phase is
\begin{align}
    \eta_\nu\equiv\lim_{z_0\to0}\frac{h_\nu^*(z_0)}{h_\nu(z_0)}  =-e^{2i\theta_\nu}\ .
    \label{eq:a4new-endpoint-phase}
\end{align}
Fixing the final massive field to zero replaces $G_{++}$ by $G_D$ and omits the opposite-branch term in \eqref{eq:a4new-response}. Using \eqref{eq:qsfi-dirichlet-green}, the resulting response and its trace completion are
\begin{align}
    r_{\nu,D}(z) &=-q_\nu\left[h_\nu^*(z)-\eta_\nu h_\nu(z)\right] +r_{\nu,{\rm an}}(z) \nonumber\\
    &=-q_\nu\sqrt\pi e^{i\theta_\nu}z^{3/2}J_\nu(z)  +r_{\nu,{\rm an}}(z)\ , \nonumber\\
    r_\nu(z)-r_{\nu,D}(z) &=-q_\nu(1+\eta_\nu)h_\nu(z) =iq_\nu\sqrt\pi\sin\theta_\nu\,z^{3/2}H^{(2)}_\nu(z)\ .
    \label{eq:a4new-dirichlet-completion}
\end{align}
The Dirichlet response has only the $z^{3/2+\nu}$ homogeneous family. The trace restores the slower $z^{3/2-\nu}$ family and changes the coefficient of the faster one.  Its effect includes both the same-branch completion and the mixed-branch contractions.  Decay of the massive field at the endpoint therefore does not justify discarding it in the full result.

For $m>3H/2$, continue the coefficients to $\nu=i\mu$ with $\mu=\sqrt{m^2/H^2-9/4}$. We then analytically continue the resulting coefficient \eqref{eq:a4new-explicit-coefficients} to $\nu = i\mu$. Since $C_+(i\mu)=C_-(i\mu)^*$, the signal is real:
\begin{align}
    \left.\Gamma^{(3,g^2;{\rm d})}_{\tilde\rho}(p,k,k)  \right|_{\rm nonanalytic}  &=\frac{2c_{\rm d}g^2k}{H^2}x^{1/2}  \operatorname{Re}\!\left[C_-(i\mu)e^{-i\mu\ln x}\right]  +\cdots\ ,  \nonumber\\
    |C_-(i\mu)|&\sim\frac{3\pi^{3/2}}4\mu^{7/2}e^{-\pi\mu}  \qquad(\mu\gg1)\ .
    \label{eq:a4new-heavy-result}
\end{align}
The omitted terms are higher powers in the same two families. For a large-mass squeezed limit, $\mu x\ll1$ is a sufficient condition for keeping the leading terms. These exponentially suppressed oscillations lie beyond any finite local inverse-mass expansion. At imaginary $\nu$, the endpoint phase $h_\nu^*(z_0)/h_\nu(z_0)$ does not have a $\tau_0 \to 0$ limit. The finite-time Dirichlet and trace pieces must therefore be added before taking $\tau_0\to0$. The real-$\nu$ separation in \eqref{eq:a4new-dirichlet-completion} is not a separate late-time prescription for either piece in this regime.

There are two useful checks on the limits.  At $\nu=1/2$, \eqref{eq:a4new-response-split} gives $r_{1/2}(z)=-z(1+iz)e^{-iz}/2$, or $R_k(\tau)=\tau K^\zeta_{+,k}(\tau)/2$, in agreement with \eqref{eq:app-conformal-linear-response}. Here $C_-=-9/2$ and $C_+=0$. The soft massive-line channel supplies $-9c_{\rm d}g^2k/(2H^2)$ at leading squeezed order; the two hard channels supply the same amount together. Their sum is the limit $-9c_{\rm d}g^2k/H^2$ of \eqref{eq:probability-conformal-g2}. At this mass the powers are integers, so the separation into analytic and nonanalytic terms no longer distinguishes the contributions. The same response in \eqref{eq:probability-lambda-diagonal-kernel} also gives the conformal self-interaction result \eqref{eq:probability-conformal-lambda}.

At $\nu=0$, the two homogeneous families overlap, and their combined limit at $\nu \to 0$ contains $x^{1/2}\ln x$. At $\nu=1$, the leading fast power coincides with the first correction to the slow family, so that correction must be included before taking $\nu \to 1$. The apparent coefficient poles cancel only after these overlapping terms have been combined. The massless endpoint $\nu=3/2$ is outside the convergence domain of the mixing integral and requires a separate treatment. Finally, the squeezed expansion describes a weakly mixed observable only while the mixing remains perturbative on the soft leg. Away from light-mass enhancements, a sufficient window at $m\sim H$ is $|g|\ll p\ll k$, as specified in Section~\ref{subsec:qsfi-squeezed-completion}.

\section{Comparison with constant proper-time mixing}
\label{app:a3-mixing}

We compare the interaction in the main text with
\begin{align}
    S_{\rm int} =\int d\tau\,d^3x\left[ ga^3\sigma F_{\rm d} -\frac{\lambda a^4}{3!}\sigma^3 \right] \ .
    \label{eq:a3-action}
\end{align}
Here the $\sigma F_{\rm d}$ mixing acquires an extra factor of $a$. All expressions in this appendix refer to this interaction. The linear mixing $g a^3 \sigma\zeta'$ now has constant cosmic-time coefficient $g$.

The heavy-field equation becomes
\begin{align}
    {\cal D}_\sigma\sigma =gaF_{\rm d}-\frac{\lambda}{2}a^2\sigma^2 \ .
    \label{eq:a3-eom}
\end{align}
The free mode functions, Green functions, and final-time trace are unchanged. The factor $a^4(\tau)$ at the $\sigma^3$ vertex also stays unchanged.

For $m^2=2H^2$, we sum the contour branches before taking $\tau_0\to0$. We keep the same perturbative orders as in the main text and assume $|g|/H\ll1$. 

The counterpart of Eq.~\eqref{eq:probability-conformal-g2} is
\begin{align}
    \Gamma^{(3,g^2;{\rm d})}_{\tilde\rho}(1,2,3) ={}&-\frac{3g^2c_{\rm d}}{H^4} \sum_{{\rm cyc}(i,j,\ell)}k_i \Bigg[ (k_j^2+k_\ell^2) \left\{ \frac{\pi^2}{3} +\operatorname{Li}_2\!\left(1-\frac{2k_i}{k_t}\right) \right\} \nonumber\\
    &\hspace{18mm} -\frac{2k_i k_jk_\ell(k_j+k_\ell)}{(k_t-2k_i)k_t} \ln\frac{k_t}{2k_i} \Bigg]  \ .
    \label{eq:a3-conformal-g2}
\end{align}
Here $\operatorname{Li}_2$ is the dilogarithm.  Unlike Eq.~\eqref{eq:probability-conformal-g2}, this result is not rational in the momenta.

The counterpart of Eq.~\eqref{eq:probability-conformal-lambda} is
\begin{align}
    \Gamma^{(3,\lambda)}_{\tilde\rho}(1,2,3) ={}&-\frac{\lambda g^3 k_1k_2k_3}{4H^7} \Bigg[ \pi^2\sum_{i=1}^3 \operatorname{Li}_2\!\left(1-\frac{2k_i}{k_t}\right) \nonumber\\
    &\quad+ \int_0^\infty\frac{dy}{y}\,e^{-y} \prod_{i=1}^3 \left\{ e^{2k_i y/k_t}E_1\!\left(\frac{2k_i y}{k_t}\right) +\gamma_E+\ln\frac{2k_i y}{k_t} \right\} \Bigg]\ .
    \label{eq:a3-conformal-lambda}
\end{align}
Here $E_1$ is the exponential integral and $\gamma_E$ is Euler's constant. The remaining integral is convergent at both endpoints and gives the complete momentum dependence at order $\lambda g^3$. In particular, this kernel is no longer momentum independent.

As in Eq.~\eqref{eq:probability-conformal-bispectrum}, the three-point function is
\begin{align}
    \left\langle \bar\zeta_{\mathbf k_1} \bar\zeta_{\mathbf k_2} \bar\zeta_{\mathbf k_3} \right\rangle_c' ={}&-\left[ \Gamma^{(3,g^2;{\rm d})}_{\tilde\rho}(1,2,3) +\Gamma^{(3,\lambda)}_{\tilde\rho}(1,2,3) \right] \prod_{a=1}^3P_\zeta(k_a)+\cdots \ .
    \label{eq:a3-conformal-bispectrum}
\end{align}
Both kernels are now homogeneous of degree three in the momenta, whereas the corresponding kernels in Eqs.~\eqref{eq:probability-conformal-g2} and~\eqref{eq:probability-conformal-lambda} have degrees one and zero. Consequently, both contributions to the three-point function scale as momentum to the power $-6$.

For $p\ll k$, their leading behavior is
\begin{align}
    \Gamma^{(3,g^2;{\rm d})}_{\tilde\rho}(p,k,k) &\simeq-\frac{2\pi^2g^2c_{\rm d}}{H^4}k^3\ , \\
    \Gamma^{(3,\lambda)}_{\tilde\rho}(p,k,k) &\simeq-\frac{\pi^4\lambda g^3}{24H^7}pk^2 \ .
    \label{eq:a3-conformal-squeezed}
\end{align}
After multiplication by the external power spectra, the first gives a constant contribution to the bispectrum divided by $P_\zeta(p)P_\zeta(k)$, while the second is proportional to $p/k$.

For $F_{\rm K}$, with the same interaction $ga^3\sigma F$, the conformal result is
\begin{align}
    \Gamma^{(3,g^2;{\rm K})}_{\tilde\rho}(1,2,3) ={}&\frac{2\pi g^2 c_{\rm K}}{H^4} \sum_{{\rm cyc}(i,j,\ell)}k_i \Bigg[ \mathbf k_j\!\cdot\!\mathbf k_\ell \left(k_j+k_\ell-k_i\ln\frac{k_t}{2k_i}\right) \nonumber\\
    &\hspace{25mm} +\frac{k_jk_\ell (k_jk_\ell-\mathbf k_j\!\cdot\!\mathbf k_\ell)} {k_t} \Bigg]\ .
    \label{eq:a3-conformal-kinetic}
\end{align}
The order-$\lambda g^3$ result is identical to that for $F_{\rm d}$ because the linear mixing is unchanged. For $p\ll k$, the kinetic kernel approaches $-2\pi g^2 c_{\rm d} pk^3/H^4$, so its bispectrum divided by $P_\zeta(p)P_\zeta(k)$ vanishes in the soft limit. Under a common rescaling of all momenta, this kernel scales with four powers of momentum, rather than the three powers found for $F_{\rm d}$. Its bispectrum therefore scales with power $-5$, rather than $-6$.

For $m\gg H$, the leading local result is
\begin{align}
    \Gamma^{(3,g^2;{\rm K})}_{\tilde\rho,\,{\rm loc}}(1,2,3)=\frac{2\pi g^2 c_{\rm K}}{H^2m^2} \sum_{{\rm cyc}(i,j,\ell)} k_i^2\,\mathbf k_j\!\cdot\!\mathbf k_\ell+\cdots .
    \label{eq:a3-local-kinetic-kernel}
\end{align}
Unlike the original $a^2$ kinetic benchmark, this contribution is nonzero. It comes from the spatial gradient term, while the leading local $\zeta'^3$ term contributes only a wavefunctional phase.

The change also affects the heavy-mass limit. For $m\gg H$ and physical frequencies below $m$, the leading local solution is $\sigma_{\rm loc}^{(0)}=gF/(am^2)$. The counterparts of the cubic terms in Eq.~\eqref{eq:heavy-local-cubic-actions} are
\begin{align}
    \left.\Delta S_{\rm loc}^{(g^2;{\rm d})}\right|_{\zeta^3} &=\frac{3g^2c_{\rm d}}{m^2} \int d\tau\,d^3x\,a^2\zeta\zeta'^2\ , \\
    \left.\Delta S_{\rm loc}^{(g^2;{\rm K})}\right|_{\zeta^3} &=\frac{g^2c_{\rm K}}{m^2} \int d\tau\,d^3x\,a^2\left[\zeta'^3-\zeta'(\nabla\zeta)^2\right]\ , \\
    \left.\Delta S_{\rm loc}^{(\lambda)}\right|_{\zeta^3} &=-\frac{\lambda g^3}{6m^6} \int d\tau\,d^3x\,a\zeta'^3\ .
    \label{eq:a3-local-cubic}
\end{align}
In particular,
\begin{align}
    \Gamma^{(3,\lambda)}_{\tilde\rho,\,{\rm loc}}(1,2,3) =\frac{4\lambda g^3}{Hm^6} \frac{(k_1k_2k_3)^2}{k_t^3}+\cdots \ .
    \label{eq:a3-local-lambda}
\end{align}
The original local $\lambda g^3$ contributes only a late-time wavefunctional phase at this order. The interaction $a\zeta'^3$ instead gives the nonzero probability kernel above, as discussed in the main text.

For $F_{\rm K}$, the spatial gradient term likewise gives a nonzero leading kernel~\eqref{eq:a3-local-kinetic-kernel}, while the local $\zeta'^3$ term contributes only a phase. The cubic boundary trace correction now generally survives as $\tau_0\to0$ at order $g^2/m^3$, so tracing over $\bar\sigma$ and fixing $\bar\sigma=0$ need not agree beyond the leading bulk order. 

The degree-three scaling of the $F_{\rm d}$ and $\lambda g^3$ conformal kernels is consistent with the linear dilation identity for a weight-zero boundary scalar. It does not by itself establish the nonlinear curvature-perturbation Ward identity, which requires the corresponding nonlinear completion of the action. The order-$g^2$ kinetic kernel instead has degree four and does not satisfy this degree-three scaling condition.

\end{appendices}

\bibliographystyle{JHEP}
\bibliography{references}

@article{Pinol:2026xnl,
    author = "Pinol, Lucas",
    title = "{New exact bispectrum shapes in multifield inflation}",
    eprint = "2607.15251",
    archivePrefix = "arXiv",
    primaryClass = "hep-th",
    month = "7",
    year = "2026"
}

@article{Wang:2026lff,
    author = "Wang, Xiangwei and Wang, Yi and Zhao, Yunke",
    title = "{Cosmological Collider Signals at Strong Mixing}",
    eprint = "2607.14891",
    archivePrefix = "arXiv",
    primaryClass = "hep-th",
    month = "7",
    year = "2026"
}

@article{Huenupi:2026abj,
    author = "Huenupi, Javier and Mu{\~n}oz, Claudio and Palma, Gonzalo A. and Sypsas, Spyros",
    title = "{Pushing the Primordial Frontier: Exact Linear Solutions in Multifield Inflation}",
    eprint = "2606.18248",
    archivePrefix = "arXiv",
    primaryClass = "astro-ph.CO",
    month = "6",
    year = "2026"
}

@article{Huenupi:2026aqc,
    author = "Huenupi, Javier and Mu{\~n}oz, Claudio and Palma, Gonzalo A. and Sypsas, Spyros",
    title = "{Pushing the Primordial Frontier: Cosmological Collider Signatures at Strong Mixing}",
    eprint = "2607.14529",
    archivePrefix = "arXiv",
    primaryClass = "astro-ph.CO",
    month = "7",
    year = "2026"
}

@article{Maldacena2002vr,
    author = "Maldacena, Juan Martin",
    title = "{Non-Gaussian features of primordial fluctuations in single field inflationary models}",
    eprint = "astro-ph/0210603",
    archivePrefix = "arXiv",
    doi = "10.1088/1126-6708/2003/05/013",
    journal = "JHEP",
    volume = "05",
    pages = "013",
    year = "2003"
}

@article{Chen2009zp,
    author = "Chen, Xingang and Wang, Yi",
    title = "{Quasi-Single Field Inflation and Non-Gaussianities}",
    eprint = "0911.3380",
    archivePrefix = "arXiv",
    primaryClass = "hep-th",
    doi = "10.1088/1475-7516/2010/04/027",
    journal = "JCAP",
    volume = "04",
    pages = "027",
    year = "2010"
}

@article{ArkaniHamed2015bza,
    author = "Arkani-Hamed, Nima and Maldacena, Juan",
    title = "{Cosmological Collider Physics}",
    eprint = "1503.08043",
    archivePrefix = "arXiv",
    primaryClass = "hep-th",
    month = "3",
    year = "2015"
}

@article{Chen2017ryl,
    author = "Chen, Xingang and Wang, Yi and Xianyu, Zhong-Zhi",
    title = "{Schwinger-Keldysh Diagrammatics for Primordial Perturbations}",
    eprint = "1703.10166",
    archivePrefix = "arXiv",
    primaryClass = "hep-th",
    doi = "10.1088/1475-7516/2017/12/006",
    journal = "JCAP",
    volume = "12",
    pages = "006",
    year = "2017"
}

@article{ArkaniHamed2018kmz,
    author = "Arkani-Hamed, Nima and Baumann, Daniel and Lee, Hayden and Pimentel, Guilherme L.",
    title = "{The Cosmological Bootstrap: Inflationary Correlators from Symmetries and Singularities}",
    eprint = "1811.00024",
    archivePrefix = "arXiv",
    primaryClass = "hep-th",
    doi = "10.1007/JHEP04(2020)105",
    journal = "JHEP",
    volume = "04",
    pages = "105",
    year = "2020"
}

@article{Baumann2022jpr,
    author = "Baumann, Daniel and Green, Daniel and Joyce, Austin and Pajer, Enrico and Pimentel, Guilherme L. and Sleight, Charlotte and Taronna, Massimo",
    title = "{Snowmass White Paper: The Cosmological Bootstrap}",
    eprint = "2203.08121",
    archivePrefix = "arXiv",
    primaryClass = "hep-th",
    doi = "10.21468/SciPostPhysCommRep.1",
    journal = "SciPost Phys. Comm. Rep.",
    volume = "2024",
    pages = "1",
    year = "2024"
}

@article{Creminelli2012ed,
    author = "Creminelli, Paolo and Nore{\~n}a, Jorge and Simonovi{\'c}, Marko",
    title = "{Conformal consistency relations for single-field inflation}",
    eprint = "1203.4595",
    archivePrefix = "arXiv",
    primaryClass = "hep-th",
    doi = "10.1088/1475-7516/2012/07/052",
    journal = "JCAP",
    volume = "07",
    pages = "052",
    year = "2012"
}

@article{Pimentel2013gza,
    author = "Pimentel, Guilherme L.",
    title = "{Inflationary Consistency Conditions from a Wavefunctional Perspective}",
    eprint = "1309.1793",
    archivePrefix = "arXiv",
    primaryClass = "hep-th",
    reportNumber = "PUPT-2452",
    doi = "10.1007/JHEP02(2014)124",
    journal = "JHEP",
    volume = "02",
    pages = "124",
    year = "2014"
}

@article{Cespedes2020Time,
    author = "C{\'e}spedes, Sebasti{\'a}n and Davis, Anne-Christine and Melville, Scott",
    title = "{On the time evolution of cosmological correlators}",
    eprint = "2009.07874",
    archivePrefix = "arXiv",
    primaryClass = "hep-th",
    doi = "10.1007/JHEP02(2021)012",
    journal = "JHEP",
    volume = "02",
    pages = "012",
    year = "2021"
}

@article{Cespedes2023IR,
    author = "C{\'e}spedes, Sebasti{\'a}n and Davis, Anne-Christine and Wang, Dong-Gang",
    title = "{On the IR divergences in de Sitter space: loops, resummation and the semi-classical wavefunction}",
    eprint = "2311.17990",
    archivePrefix = "arXiv",
    primaryClass = "hep-th",
    doi = "10.1007/JHEP04(2024)004",
    journal = "JHEP",
    volume = "04",
    pages = "004",
    year = "2024"
}

@article{Pimentel2022Boostless,
    author = "Pimentel, Guilherme L. and Wang, Dong-Gang",
    title = "{Boostless cosmological collider bootstrap}",
    eprint = "2205.00013",
    archivePrefix = "arXiv",
    primaryClass = "hep-th",
    doi = "10.1007/JHEP10(2022)177",
    journal = "JHEP",
    volume = "10",
    pages = "177",
    year = "2022"
}

@article{Palma2026Map,
    author = "Palma, Gonzalo A.",
    title = "{From the wavefunction of the Universe to in-in-correlators: a perturbative map to all orders}",
    eprint = "2601.00992",
    archivePrefix = "arXiv",
    primaryClass = "hep-th",
    doi = "10.1007/JHEP06(2026)179",
    journal = "JHEP",
    volume = "06",
    pages = "179",
    year = "2026"
}

@article{Goodhew2020hob,
    author = "Goodhew, Harry and Jazayeri, Sadra and Pajer, Enrico",
    title = "{The Cosmological Optical Theorem}",
    eprint = "2009.02898",
    archivePrefix = "arXiv",
    primaryClass = "hep-th",
    doi = "10.1088/1475-7516/2021/04/021",
    journal = "JCAP",
    volume = "04",
    pages = "021",
    year = "2021"
}

@article{Braglia2024Derivatives,
    author = "Braglia, Matteo and Pinol, Lucas",
    title = "{No time to derive: unraveling total time derivatives in in-in perturbation theory}",
    eprint = "2403.14558",
    archivePrefix = "arXiv",
    primaryClass = "astro-ph.CO",
    doi = "10.1007/JHEP08(2024)068",
    journal = "JHEP",
    volume = "08",
    pages = "068",
    year = "2024"
}

@article{Chen2009we,
    author = "Chen, Xingang and Wang, Yi",
    title = "{Large non-Gaussianities with Intermediate Shapes from Quasi-Single Field Inflation}",
    eprint = "0909.0496",
    archivePrefix = "arXiv",
    primaryClass = "astro-ph.CO",
    reportNumber = "MIT-CTP-4071",
    doi = "10.1103/PhysRevD.81.063511",
    journal = "Phys. Rev. D",
    volume = "81",
    pages = "063511",
    year = "2010"
}

@article{Werth2023Flow,
    author = "Werth, Denis and Pinol, Lucas and Renaux-Petel, S{\'e}bastien",
    title = "{Cosmological Flow of Primordial Correlators}",
    eprint = "2302.00655",
    archivePrefix = "arXiv",
    primaryClass = "hep-th",
    doi = "10.1103/PhysRevLett.133.141002",
    journal = "Phys. Rev. Lett.",
    volume = "133",
    number = "14",
    pages = "141002",
    year = "2024"
}

@article{Cespedes2012hu,
    author = "Cespedes, Sebastian and Atal, Vicente and Palma, Gonzalo A.",
    title = "{On the importance of heavy fields during inflation}",
    eprint = "1201.4848",
    archivePrefix = "arXiv",
    primaryClass = "hep-th",
    doi = "10.1088/1475-7516/2012/05/008",
    journal = "JCAP",
    volume = "05",
    pages = "008",
    year = "2012"
}

@article{Achucarro2012sm,
    author = "Achucarro, Ana and Gong, Jinn-Ouk and Hardeman, Sjoerd and Palma, Gonzalo A. and Patil, Subodh P.",
    title = "{Effective theories of single field inflation when heavy fields matter}",
    eprint = "1201.6342",
    archivePrefix = "arXiv",
    primaryClass = "hep-th",
    reportNumber = "CERN-PH-TH-2011-222, CPHT-RR-055.0711, LPTENS-11-27",
    doi = "10.1007/JHEP05(2012)066",
    journal = "JHEP",
    volume = "05",
    pages = "066",
    year = "2012"
}

@article{Cheung2007st,
    author = "Cheung, Clifford and Creminelli, Paolo and Fitzpatrick, A. Liam and Kaplan, Jared and Senatore, Leonardo",
    title = "{The Effective Field Theory of Inflation}",
    eprint = "0709.0293",
    archivePrefix = "arXiv",
    primaryClass = "hep-th",
    reportNumber = "IC-2007-032",
    doi = "10.1088/1126-6708/2008/03/014",
    journal = "JHEP",
    volume = "03",
    pages = "014",
    year = "2008"
}

@article{Weinberg2005vy,
    author = "Weinberg, Steven",
    title = "{Quantum contributions to cosmological correlations}",
    eprint = "hep-th/0506236",
    archivePrefix = "arXiv",
    reportNumber = "UTTG-01-05",
    doi = "10.1103/PhysRevD.72.043514",
    journal = "Phys. Rev. D",
    volume = "72",
    pages = "043514",
    year = "2005"
}

@article{Qin2023Seeds,
    author = "Qin, Zhehan and Xianyu, Zhong-Zhi",
    title = "{Closed-form formulae for inflation correlators}",
    eprint = "2301.07047",
    archivePrefix = "arXiv",
    primaryClass = "hep-th",
    doi = "10.1007/JHEP07(2023)001",
    journal = "JHEP",
    volume = "07",
    pages = "001",
    year = "2023"
}

@article{Cespedes2025Quartic,
    author = "Cespedes, Sebastian and Qin, Zhehan and Wang, Dong-Gang",
    title = "{{\ensuremath{\lambda}}{\ensuremath{\phi}}$^{4}$ as an effective theory in de Sitter}",
    eprint = "2510.25826",
    archivePrefix = "arXiv",
    primaryClass = "hep-th",
    reportNumber = "Imperial-TP-2025-SCC-3",
    doi = "10.1007/JHEP05(2026)143",
    journal = "JHEP",
    volume = "05",
    pages = "143",
    year = "2026"
}

@article{GreenSun2025,
    author = "Green, Daniel and Sun, Guanhao",
    title = "{Effective field theory and in-in correlators}",
    eprint = "2412.02739",
    archivePrefix = "arXiv",
    primaryClass = "hep-th",
    doi = "10.1007/JHEP04(2025)166",
    journal = "JHEP",
    volume = "04",
    pages = "166",
    year = "2025"
}

@article{AguiSalcedo2024Open,
    author = "Agui Salcedo, Santiago and Colas, Thomas and Pajer, Enrico",
    title = "{The open effective field theory of inflation}",
    eprint = "2404.15416",
    archivePrefix = "arXiv",
    primaryClass = "hep-th",
    doi = "10.1007/JHEP10(2024)248",
    journal = "JHEP",
    volume = "10",
    pages = "248",
    year = "2024"
}

@article{ColasQinTong2026Collider,
    author = "Colas, Thomas and Qin, Zhehan and Tong, Xi",
    title = "{Open effective field theory and the physics of cosmological collider signals}",
    eprint = "2512.07941",
    archivePrefix = "arXiv",
    primaryClass = "hep-th",
    doi = "10.1007/JHEP08(2026)032",
    journal = "JHEP",
    volume = "08",
    pages = "032",
    year = "2026"
}

@article{Boyanovsky2015Effective,
    author = "Boyanovsky, D.",
    title = "{Effective field theory during inflation: Reduced density matrix and its quantum master equation}",
    eprint = "1506.07395",
    archivePrefix = "arXiv",
    primaryClass = "astro-ph.CO",
    doi = "10.1103/PhysRevD.92.023527",
    journal = "Phys. Rev. D",
    volume = "92",
    number = "2",
    pages = "023527",
    year = "2015"
}

@article{Boyanovsky2016Stochastic,
    author = "Boyanovsky, D.",
    title = "{Effective field theory during inflation. II. Stochastic dynamics and power spectrum suppression}",
    eprint = "1511.06649",
    archivePrefix = "arXiv",
    primaryClass = "astro-ph.CO",
    doi = "10.1103/PhysRevD.93.043501",
    journal = "Phys. Rev. D",
    volume = "93",
    pages = "043501",
    year = "2016"
}

@article{Colas2022Open,
    author = "Colas, Thomas and Grain, Julien and Vennin, Vincent",
    title = "{Benchmarking the cosmological master equations}",
    eprint = "2209.01929",
    archivePrefix = "arXiv",
    primaryClass = "hep-th",
    doi = "10.1140/epjc/s10052-022-11047-9",
    journal = "Eur. Phys. J. C",
    volume = "82",
    number = "12",
    pages = "1085",
    year = "2022"
}

@inproceedings{Colas2025Lectures,
    author = "Colas, Thomas",
    title = "{Lectures on Open Effective Field Theories}",
    eprint = "2510.00140",
    archivePrefix = "arXiv",
    primaryClass = "hep-th",
    month = "9",
    year = "2025"
}

@article{Pajer2026Lectures,
    author = "Pajer, Enrico",
    title = "{Lectures on Open Systems and Cosmology:~Master Equation, Schwinger{\textendash}Keldysh and Open Effective Field Theories}",
    eprint = "2607.14351",
    archivePrefix = "arXiv",
    primaryClass = "hep-th",
    month = "7",
    year = "2026"
}

@article{CespedesColas2026,
    author = "C{\'e}spedes, Sebasti{\'a}n and Colas, Thomas",
    title = "{Stochastic inflation from a non-equilibrium renormalization group}",
    eprint = "2605.11096",
    archivePrefix = "arXiv",
    primaryClass = "hep-th",
    month = "5",
    year = "2026"
}

@article{FeynmanVernon1963,
    author = "Feynman, R. P. and Vernon, Jr., F. L.",
    editor = "Brown, L. M.",
    title = "{The Theory of a general quantum system interacting with a linear dissipative system}",
    doi = "10.1016/0003-4916(63)90068-X",
    journal = "Annals Phys.",
    volume = "24",
    pages = "118--173",
    year = "1963"
}

@article{Burgess2015Open,
    author = "Burgess, C. P. and Holman, R. and Tasinato, G. and Williams, M.",
    title = "{EFT Beyond the Horizon: Stochastic Inflation and How Primordial Quantum Fluctuations Go Classical}",
    eprint = "1408.5002",
    archivePrefix = "arXiv",
    primaryClass = "hep-th",
    reportNumber = "CERN-PH-TH-2014-142",
    doi = "10.1007/JHEP03(2015)090",
    journal = "JHEP",
    volume = "03",
    pages = "090",
    year = "2015"
}

@article{Schwinger1961Brownian,
    author = "Schwinger, Julian S.",
    title = "{Brownian motion of a quantum oscillator}",
    doi = "10.1063/1.1703727",
    journal = "J. Math. Phys.",
    volume = "2",
    pages = "407--432",
    year = "1961"
}

@article{Keldysh1965,
    author = "Keldysh, L. V.",
    title = "{Diagram Technique for Nonequilibrium Processes}",
    doi = "10.1142/9789811279461_0007",
    journal = "Sov. Phys. JETP",
    volume = "20",
    pages = "1018--1026",
    year = "1965"
}

@article{Jordan1986Causal,
    author = "Jordan, R. D.",
    title = "{Effective Field Equations for Expectation Values}",
    doi = "10.1103/PhysRevD.33.444",
    journal = "Phys. Rev. D",
    volume = "33",
    pages = "444--454",
    year = "1986"
}

@article{CalzettaHu1987CTP,
    author = "Calzetta, E. and Hu, B. L.",
    title = "{Closed Time Path Functional Formalism in Curved Space-Time: Application to Cosmological Back Reaction Problems}",
    reportNumber = "MdDP-PP-86-189",
    doi = "10.1103/PhysRevD.35.495",
    journal = "Phys. Rev. D",
    volume = "35",
    pages = "495",
    year = "1987"
}

@article{CrossleyGloriosoLiu2017,
    author = "Crossley, Michael and Glorioso, Paolo and Liu, Hong",
    title = "{Effective field theory of dissipative fluids}",
    eprint = "1511.03646",
    archivePrefix = "arXiv",
    primaryClass = "hep-th",
    reportNumber = "MIT-CTP-4734",
    doi = "10.1007/JHEP09(2017)095",
    journal = "JHEP",
    volume = "09",
    pages = "095",
    year = "2017"
}

@article{NoumiYamaguchiYokoyama2013,
    author = "Noumi, Toshifumi and Yamaguchi, Masahide and Yokoyama, Daisuke",
    title = "{Effective field theory approach to quasi-single field inflation and effects of heavy fields}",
    eprint = "1211.1624",
    archivePrefix = "arXiv",
    primaryClass = "hep-th",
    reportNumber = "UT-KOMABA-12-9, TIT-HEP-625",
    doi = "10.1007/JHEP06(2013)051",
    journal = "JHEP",
    volume = "06",
    pages = "051",
    year = "2013"
}

@article{Achucarro2012Decoupling,
    author = "Achucarro, Ana and Atal, Vicente and Cespedes, Sebastian and Gong, Jinn-Ouk and Palma, Gonzalo A. and Patil, Subodh P.",
    title = "{Heavy fields, reduced speeds of sound and decoupling during inflation}",
    eprint = "1205.0710",
    archivePrefix = "arXiv",
    primaryClass = "hep-th",
    reportNumber = "CERN-PH-TH-2012-097, CPHT-RR-020.0412",
    doi = "10.1103/PhysRevD.86.121301",
    journal = "Phys. Rev. D",
    volume = "86",
    pages = "121301",
    year = "2012"
}

@article{GongPiSasaki2013Equilateral,
    author = "Gong, Jinn-Ouk and Pi, Shi and Sasaki, Misao",
    title = "{Equilateral non-Gaussianity from heavy fields}",
    eprint = "1306.3691",
    archivePrefix = "arXiv",
    primaryClass = "hep-th",
    doi = "10.1088/1475-7516/2013/11/043",
    journal = "JCAP",
    volume = "11",
    pages = "043",
    year = "2013"
}

@article{TongWangZhou2018Warm,
    author = "Tong, Xi and Wang, Yi and Zhou, Siyi",
    title = "{Unsuppressed primordial standard clocks in warm quasi-single field inflation}",
    eprint = "1801.05688",
    archivePrefix = "arXiv",
    primaryClass = "hep-th",
    doi = "10.1088/1475-7516/2018/06/013",
    journal = "JCAP",
    volume = "06",
    pages = "013",
    year = "2018"
}

@article{GoodhewJazayeriLeePajer2021,
    author = "Goodhew, Harry and Jazayeri, Sadra and Lee, Mang Hei Gordon and Pajer, Enrico",
    title = "{Cutting cosmological correlators}",
    eprint = "2104.06587",
    archivePrefix = "arXiv",
    primaryClass = "hep-th",
    doi = "10.1088/1475-7516/2021/08/003",
    journal = "JCAP",
    volume = "08",
    pages = "003",
    year = "2021"
}

@article{AguiSalcedo2023Analytic,
    author = "Agui Salcedo, Santiago and Lee, Mang Hei Gordon and Melville, Scott and Pajer, Enrico",
    title = "{The Analytic Wavefunction}",
    eprint = "2212.08009",
    archivePrefix = "arXiv",
    primaryClass = "hep-th",
    doi = "10.1007/JHEP06(2023)020",
    journal = "JHEP",
    volume = "06",
    pages = "020",
    year = "2023"
}

@article{DuasoPueyo2026Asymptotic,
    author = "Duaso Pueyo, Carlos and Goodhew, Harry and McCulloch, Ciaran and Pajer, Enrico",
    title = "{On the asymptotic nature of cosmological effective theories}",
    eprint = "2505.17820",
    archivePrefix = "arXiv",
    primaryClass = "hep-th",
    doi = "10.1007/JHEP01(2026)009",
    journal = "JHEP",
    volume = "01",
    pages = "009",
    year = "2026"
}

@article{LopezNacir2012Dissipative,
    author = "Lopez Nacir, Diana and Porto, Rafael A. and Senatore, Leonardo and Zaldarriaga, Matias",
    title = "{Dissipative effects in the Effective Field Theory of Inflation}",
    eprint = "1109.4192",
    archivePrefix = "arXiv",
    primaryClass = "hep-th",
    reportNumber = "SLAC-PUB-14995",
    doi = "10.1007/JHEP01(2012)075",
    journal = "JHEP",
    volume = "01",
    pages = "075",
    year = "2012"
}

@article{Galley2013Nonconservative,
    author = "Galley, Chad R.",
    title = "{Classical Mechanics of Nonconservative Systems}",
    eprint = "1210.2745",
    archivePrefix = "arXiv",
    primaryClass = "gr-qc",
    doi = "10.1103/PhysRevLett.110.174301",
    journal = "Phys. Rev. Lett.",
    volume = "110",
    number = "17",
    pages = "174301",
    year = "2013"
}

@article{GalleyTsangStein2014,
    author = "Galley, Chad R. and Tsang, David and Stein, Leo C.",
    title = "{The principle of stationary nonconservative action for classical mechanics and field theories}",
    eprint = "1412.3082",
    archivePrefix = "arXiv",
    primaryClass = "math-ph",
    month = "12",
    year = "2014"
}

@article{Sohn2024Collider,
    author = "Sohn, Wuhyun and Wang, Dong-Gang and Fergusson, James R. and Shellard, E. P. S.",
    title = "{Searching for cosmological collider in the Planck CMB data}",
    eprint = "2404.07203",
    archivePrefix = "arXiv",
    primaryClass = "astro-ph.CO",
    doi = "10.1088/1475-7516/2024/09/016",
    journal = "JCAP",
    volume = "09",
    pages = "016",
    year = "2024"
}

@article{Suman2025ColliderII,
    author = "Suman, Petar and Wang, Dong-Gang and Sohn, Wuhyun and Fergusson, James R. and Shellard, E. P. S.",
    title = "{Searching for Cosmological Collider in the Planck CMB Data II: collider templates and Modal analysis}",
    eprint = "2512.22085",
    archivePrefix = "arXiv",
    primaryClass = "astro-ph.CO",
    month = "12",
    year = "2025"
}

@article{Kumar2026ScalarSearch,
    author = "Kumar, Soubhik and Lu, Qianshu and Xianyu, Zhong-Zhi and Zhang, Yisong",
    title = "{Scalars at the Cosmological Collider: Full Shapes of Tree Diagrams and Bispectrum Searches using Planck Data}",
    eprint = "2604.07434",
    archivePrefix = "arXiv",
    primaryClass = "hep-ph",
    month = "4",
    year = "2026"
}

@article{Philcox2026ScalarCMB,
    author = "Philcox, Oliver H. E.",
    title = "{Dissecting the Scalar Cosmological Collider with the Cosmic Microwave Background}",
    eprint = "2607.18369",
    archivePrefix = "arXiv",
    primaryClass = "astro-ph.CO",
    month = "7",
    year = "2026"
}

@article{Cabass2025BOSS,
    author = "Cabass, Giovanni and Philcox, Oliver H. E. and Ivanov, Mikhail M. and Akitsu, Kazuyuki and Chen, Shi-Fan and Simonovi{\'c}, Marko and Zaldarriaga, Matias",
    title = "{BOSS constraints on massive particles during inflation: The cosmological collider in action}",
    eprint = "2404.01894",
    archivePrefix = "arXiv",
    primaryClass = "astro-ph.CO",
    reportNumber = "RBI-ThPhys-2024-21, MIT-CTP/5698",
    doi = "10.1103/PhysRevD.111.063510",
    journal = "Phys. Rev. D",
    volume = "111",
    number = "6",
    pages = "063510",
    year = "2025"
}

@article{Sefusatti2012QSF,
    author = "Sefusatti, Emiliano and Fergusson, James R. and Chen, Xingang and Shellard, E. P. S.",
    title = "{Effects and Detectability of Quasi-Single Field Inflation in the Large-Scale Structure and Cosmic Microwave Background}",
    eprint = "1204.6318",
    archivePrefix = "arXiv",
    primaryClass = "astro-ph.CO",
    doi = "10.1088/1475-7516/2012/08/033",
    journal = "JCAP",
    volume = "08",
    pages = "033",
    year = "2012"
}

@article{Dizgah2018GalaxyBispectrum,
    author = "Moradinezhad Dizgah, Azadeh and Lee, Hayden and Mu{\~n}oz, Julian B. and Dvorkin, Cora",
    title = "{Galaxy Bispectrum from Massive Spinning Particles}",
    eprint = "1801.07265",
    archivePrefix = "arXiv",
    primaryClass = "astro-ph.CO",
    doi = "10.1088/1475-7516/2018/05/013",
    journal = "JCAP",
    volume = "05",
    pages = "013",
    year = "2018"
}

@article{Dizgah2018ScaleBias,
    author = "Moradinezhad Dizgah, Azadeh and Dvorkin, Cora",
    title = "{Scale-Dependent Galaxy Bias from Massive Particles with Spin during Inflation}",
    eprint = "1708.06473",
    archivePrefix = "arXiv",
    primaryClass = "astro-ph.CO",
    doi = "10.1088/1475-7516/2018/01/010",
    journal = "JCAP",
    volume = "01",
    pages = "010",
    year = "2018"
}

@article{Meerburg2017Prospects,
    author = {Meerburg, P. Daniel and M{\"u}nchmeyer, Moritz and Mu{\~n}oz, Julian B. and Chen, Xingang},
    title = "{Prospects for Cosmological Collider Physics}",
    eprint = "1610.06559",
    archivePrefix = "arXiv",
    primaryClass = "astro-ph.CO",
    doi = "10.1088/1475-7516/2017/03/050",
    journal = "JCAP",
    volume = "03",
    pages = "050",
    year = "2017"
}

@article{Dizgah2018LongLived,
    author = "Moradinezhad Dizgah, Azadeh and Franciolini, Gabriele and Kehagias, Alex and Riotto, Antonio",
    title = "{Constraints on long-lived, higher-spin particles from galaxy bispectrum}",
    eprint = "1805.10247",
    archivePrefix = "arXiv",
    primaryClass = "astro-ph.CO",
    doi = "10.1103/PhysRevD.98.063520",
    journal = "Phys. Rev. D",
    volume = "98",
    number = "6",
    pages = "063520",
    year = "2018"
}

@article{Assassi2015Bias,
    author = "Assassi, Valentin and Baumann, Daniel and Schmidt, Fabian",
    title = "{Galaxy Bias and Primordial Non-Gaussianity}",
    eprint = "1510.03723",
    archivePrefix = "arXiv",
    primaryClass = "astro-ph.CO",
    doi = "10.1088/1475-7516/2015/12/043",
    journal = "JCAP",
    volume = "12",
    pages = "043",
    year = "2015"
}

@article{Planck2018NG,
    author = "Akrami, Y. and others",
    collaboration = "Planck",
    title = "{Planck 2018 results. IX. Constraints on primordial non-Gaussianity}",
    eprint = "1905.05697",
    archivePrefix = "arXiv",
    primaryClass = "astro-ph.CO",
    doi = "10.1051/0004-6361/201935891",
    journal = "Astron. Astrophys.",
    volume = "641",
    pages = "A9",
    year = "2020"
}

@article{Iyer2018PartialEFT,
    author = "Iyer, Aditya Varna and Pi, Shi and Wang, Yi and Wang, Ziwei and Zhou, Siyi",
    title = "{Strongly Coupled Quasi-Single Field Inflation}",
    eprint = "1710.03054",
    archivePrefix = "arXiv",
    primaryClass = "hep-th",
    doi = "10.1088/1475-7516/2018/01/041",
    journal = "JCAP",
    volume = "01",
    pages = "041",
    year = "2018"
}

@article{Chen2016QuantumClocks,
    author = "Chen, Xingang and Namjoo, Mohammad Hossein and Wang, Yi",
    title = "{Quantum Primordial Standard Clocks}",
    eprint = "1509.03930",
    archivePrefix = "arXiv",
    primaryClass = "astro-ph.CO",
    doi = "10.1088/1475-7516/2016/02/013",
    journal = "JCAP",
    volume = "02",
    pages = "013",
    year = "2016"
}

@article{Hongo2019Time,
    author = "Hongo, Masaru and Kim, Suro and Noumi, Toshifumi and Ota, Atsuhisa",
    title = "{Effective field theory of time-translational symmetry breaking in nonequilibrium open system}",
    eprint = "1805.06240",
    archivePrefix = "arXiv",
    primaryClass = "hep-th",
    reportNumber = "RIKEN-iTHEMS-Report-18, KOBE-COSMO-18-05, RIKEN-ITHEMS-REPORT-18",
    doi = "10.1007/JHEP02(2019)131",
    journal = "JHEP",
    volume = "02",
    pages = "131",
    year = "2019"
}

@article{Chen2012LargeMass,
    author = "Chen, Xingang and Wang, Yi",
    title = "{Quasi-Single Field Inflation with Large Mass}",
    eprint = "1205.0160",
    archivePrefix = "arXiv",
    primaryClass = "hep-th",
    doi = "10.1088/1475-7516/2012/09/021",
    journal = "JCAP",
    volume = "09",
    pages = "021",
    year = "2012"
}

@article{Green:2020ebl,
    author = "Green, Daniel and Pajer, Enrico",
    title = "{On the Symmetries of Cosmological Perturbations}",
    eprint = "2004.09587",
    archivePrefix = "arXiv",
    primaryClass = "hep-th",
    doi = "10.1088/1475-7516/2020/09/032",
    journal = "JCAP",
    volume = "09",
    pages = "032",
    year = "2020"
}

@article{Assassi:2012zq,
    author = "Assassi, Valentin and Baumann, Daniel and Green, Daniel",
    title = "{On Soft Limits of Inflationary Correlation Functions}",
    eprint = "1204.4207",
    archivePrefix = "arXiv",
    primaryClass = "hep-th",
    doi = "10.1088/1475-7516/2012/11/047",
    journal = "JCAP",
    volume = "11",
    pages = "047",
    year = "2012"
}

@article{Cohen:2020php,
    author = "Cohen, Timothy and Green, Daniel",
    title = "{Soft de Sitter Effective Theory}",
    eprint = "2007.03693",
    archivePrefix = "arXiv",
    primaryClass = "hep-th",
    doi = "10.1007/JHEP12(2020)041",
    journal = "JHEP",
    volume = "12",
    pages = "041",
    year = "2020"
}

@article{Green:2025hmo,
    author = "Green, Daniel and Gupta, Kshitij",
    title = "{Quantum walks and exact RG in de Sitter space}",
    eprint = "2512.13842",
    archivePrefix = "arXiv",
    primaryClass = "hep-th",
    doi = "10.1103/jz1v-8lfy",
    journal = "Phys. Rev. D",
    volume = "114",
    number = "6",
    pages = "063525",
    year = "2026"
}

@article{Burgess:2022nwu,
    author = "Burgess, C. P. and Holman, R. and Kaplanek, Greg and Martin, Jerome and Vennin, Vincent",
    title = "{Minimal decoherence from inflation}",
    eprint = "2211.11046",
    archivePrefix = "arXiv",
    primaryClass = "hep-th",
    reportNumber = "CERN-TH-2022-174; Imperial/TP/2022/GK/02",
    doi = "10.1088/1475-7516/2023/07/022",
    journal = "JCAP",
    volume = "07",
    pages = "022",
    year = "2023"
}

@article{Burgess:2024eng,
    author = "Burgess, C. P. and Colas, Thomas and Holman, R. and Kaplanek, Greg and Vennin, Vincent",
    title = "{Cosmic purity lost: perturbative and resummed late-time inflationary decoherence}",
    eprint = "2403.12240",
    archivePrefix = "arXiv",
    primaryClass = "gr-qc",
    doi = "10.1088/1475-7516/2024/08/042",
    journal = "JCAP",
    volume = "08",
    pages = "042",
    year = "2024"
}

@article{Xianyu:2023ytd,
    author = "Xianyu, Zhong-Zhi and Zang, Jiaju",
    title = "{Inflation correlators with multiple massive exchanges}",
    eprint = "2309.10849",
    archivePrefix = "arXiv",
    primaryClass = "hep-th",
    doi = "10.1007/JHEP03(2024)070",
    journal = "JHEP",
    volume = "03",
    pages = "070",
    year = "2024"
}

@article{Qin:2025xct,
    author = "Qin, Zhehan and Renaux-Petel, S{\'e}bastien and Tong, Xi and Werth, Denis and Zhu, Yuhang",
    title = "{The exact and approximate tales of boost-breaking cosmological correlators}",
    eprint = "2506.01555",
    archivePrefix = "arXiv",
    primaryClass = "hep-th",
    doi = "10.1088/1475-7516/2025/09/058",
    journal = "JCAP",
    volume = "09",
    pages = "058",
    year = "2025"
}

@article{Melville:2021lst,
    author = "Melville, Scott and Pajer, Enrico",
    title = "{Cosmological Cutting Rules}",
    eprint = "2103.09832",
    archivePrefix = "arXiv",
    primaryClass = "hep-th",
    doi = "10.1007/JHEP05(2021)249",
    journal = "JHEP",
    volume = "05",
    pages = "249",
    year = "2021"
}

@article{Jazayeri:2021fvk,
    author = "Jazayeri, Sadra and Pajer, Enrico and Stefanyszyn, David",
    title = "{From locality and unitarity to cosmological correlators}",
    eprint = "2103.08649",
    archivePrefix = "arXiv",
    primaryClass = "hep-th",
    doi = "10.1007/JHEP10(2021)065",
    journal = "JHEP",
    volume = "10",
    pages = "065",
    year = "2021"
}

@article{Stefanyszyn:2024msm,
    author = "Stefanyszyn, David and Tong, Xi and Zhu, Yuhang",
    title = "{There and Back Again: Mapping and Factorizing Cosmological Observables}",
    eprint = "2406.00099",
    archivePrefix = "arXiv",
    primaryClass = "hep-th",
    doi = "10.1103/PhysRevLett.133.221501",
    journal = "Phys. Rev. Lett.",
    volume = "133",
    number = "22",
    pages = "221501",
    year = "2024"
}

@article{Kawaguchi:2024lsw,
    author = "Kawaguchi, Ryodai and Tsujikawa, Shinji and Yamada, Yusuke",
    title = "{Roles of boundary and equation-of-motion terms in cosmological correlation functions}",
    eprint = "2403.16022",
    archivePrefix = "arXiv",
    primaryClass = "hep-th",
    reportNumber = "WUCG-24-03",
    doi = "10.1016/j.physletb.2024.138962",
    journal = "Phys. Lett. B",
    volume = "856",
    pages = "138962",
    year = "2024"
}

@article{Brahma:2024yor,
    author = "Brahma, Suddhasattwa and Calder{\'o}n-Figueroa, Jaime and Luo, Xiancong",
    title = "{Time-convolutionless cosmological master equations: late-time resummations and decoherence for non-local kernels}",
    eprint = "2407.12091",
    archivePrefix = "arXiv",
    primaryClass = "hep-th",
    doi = "10.1088/1475-7516/2025/08/019",
    journal = "JCAP",
    volume = "08",
    pages = "019",
    year = "2025"
}

@article{Pajer:2020wxk,
    author = "Pajer, Enrico",
    title = "{Building a Boostless Bootstrap for the Bispectrum}",
    eprint = "2010.12818",
    archivePrefix = "arXiv",
    primaryClass = "hep-th",
    doi = "10.1088/1475-7516/2021/01/023",
    journal = "JCAP",
    volume = "01",
    pages = "023",
    year = "2021"
}

@article{AguiSalcedo:2023nds,
    author = "Agui Salcedo, Santiago and Melville, Scott",
    title = "{The cosmological tree theorem}",
    eprint = "2308.00680",
    archivePrefix = "arXiv",
    primaryClass = "hep-th",
    doi = "10.1007/JHEP12(2023)076",
    journal = "JHEP",
    volume = "12",
    pages = "076",
    year = "2023"
}

@article{Arkani-Hamed:2025mce,
    author = "Arkani-Hamed, Nima and Glew, Ross and Vaz{\~a}o, Francisco",
    title = "{Correlators Are Simpler than Wave Functions}",
    eprint = "2512.23795",
    archivePrefix = "arXiv",
    primaryClass = "hep-th",
    doi = "10.1103/sgm7-181s",
    journal = "Phys. Rev. Lett.",
    volume = "137",
    number = "6",
    pages = "061601",
    year = "2026"
}

@article{Chowdhury:2026dwm,
    author = "Chowdhury, Chandramouli and He, Song and Su, Yong-Xiang and Yang, Dongyu",
    title = "{On the simplicity of de Sitter correlators}",
    eprint = "2604.26421",
    archivePrefix = "arXiv",
    primaryClass = "hep-th",
    doi = "10.1007/JHEP08(2026)205",
    journal = "JHEP",
    volume = "08",
    pages = "205",
    year = "2026"
}

@article{Qin:2024gtr,
    author = "Qin, Zhehan",
    title = "{Cosmological correlators at the loop level}",
    eprint = "2411.13636",
    archivePrefix = "arXiv",
    primaryClass = "hep-th",
    doi = "10.1007/JHEP03(2025)051",
    journal = "JHEP",
    volume = "03",
    pages = "051",
    year = "2025"
}

@article{Goodhew:2024eup,
    author = "Goodhew, Harry and Thavanesan, Ayngaran and Wall, Aron C.",
    title = "{The cosmological CPT theorem}",
    eprint = "2408.17406",
    archivePrefix = "arXiv",
    primaryClass = "hep-th",
    doi = "10.21468/SciPostPhys.20.2.049",
    journal = "SciPost Phys.",
    volume = "20",
    number = "2",
    pages = "049",
    year = "2026"
}

@article{Thavanesan:2025kyc,
    author = "Thavanesan, Ayngaran",
    title = "{No-go Theorem for Cosmological Parity Violation}",
    eprint = "2501.06383",
    archivePrefix = "arXiv",
    primaryClass = "hep-th",
    month = "1",
    year = "2025"
}

@article{Aoki:2024uyi,
    author = "Aoki, Shuntaro and Pinol, Lucas and Sano, Fumiya and Yamaguchi, Masahide and Zhu, Yuhang",
    title = "{Cosmological correlators with double massive exchanges: bootstrap equation and phenomenology}",
    eprint = "2404.09547",
    archivePrefix = "arXiv",
    primaryClass = "hep-th",
    doi = "10.1007/JHEP09(2024)176",
    journal = "JHEP",
    volume = "09",
    pages = "176",
    year = "2024"
}

@article{Aoki:2023wdc,
    author = "Aoki, Shuntaro and Noumi, Toshifumi and Sano, Fumiya and Yamaguchi, Masahide",
    title = "{Analytic formulae for inflationary correlators with dynamical mass}",
    eprint = "2312.09642",
    archivePrefix = "arXiv",
    primaryClass = "hep-th",
    reportNumber = "CTPU-PTC-23-47, UT-Komaba/23-13",
    doi = "10.1007/JHEP03(2024)073",
    journal = "JHEP",
    volume = "03",
    pages = "073",
    year = "2024"
}

@article{An:2017hlx,
    author = "An, Haipeng and McAneny, Michael and Ridgway, Alexander K. and Wise, Mark B.",
    title = "{Quasi Single Field Inflation in the non-perturbative regime}",
    eprint = "1706.09971",
    archivePrefix = "arXiv",
    primaryClass = "hep-ph",
    doi = "10.1007/JHEP06(2018)105",
    journal = "JHEP",
    volume = "06",
    pages = "105",
    year = "2018"
}

@article{Tong:2017iat,
    author = "Tong, Xi and Wang, Yi and Zhou, Siyi",
    title = "{On the Effective Field Theory for Quasi-Single Field Inflation}",
    eprint = "1708.01709",
    archivePrefix = "arXiv",
    primaryClass = "astro-ph.CO",
    doi = "10.1088/1475-7516/2017/11/045",
    journal = "JCAP",
    volume = "11",
    pages = "045",
    year = "2017"
}

@article{Agarwal:2012mq,
    author = "Agarwal, Nishant and Holman, R. and Tolley, Andrew J. and Lin, Jennifer",
    title = "{Effective field theory and non-Gaussianity from general inflationary states}",
    eprint = "1212.1172",
    archivePrefix = "arXiv",
    primaryClass = "hep-th",
    doi = "10.1007/JHEP05(2013)085",
    journal = "JHEP",
    volume = "05",
    pages = "085",
    year = "2013"
}

@article{Yin:2023jlv,
    author = "Yin, Yuan",
    title = "{Cosmological collider signal from non-Bunch-Davies initial states}",
    eprint = "2309.05244",
    archivePrefix = "arXiv",
    primaryClass = "hep-ph",
    doi = "10.1103/PhysRevD.109.043535",
    journal = "Phys. Rev. D",
    volume = "109",
    number = "4",
    pages = "043535",
    year = "2024"
}

@article{Chopping:2024oiu,
    author = "Chopping, Alistair J. and Sleight, Charlotte and Taronna, Massimo",
    title = "{Cosmological correlators for Bogoliubov initial states}",
    eprint = "2407.16652",
    archivePrefix = "arXiv",
    primaryClass = "hep-th",
    doi = "10.1007/JHEP09(2024)152",
    journal = "JHEP",
    volume = "09",
    pages = "152",
    year = "2024"
}

@article{Ghosh:2025pxn,
    author = "Ghosh, Diptimoy and Ullah, Farman",
    title = "{Cosmological cutting rules for Bogoliubov initial states: any mass and spin}",
    eprint = "2502.05630",
    archivePrefix = "arXiv",
    primaryClass = "hep-th",
    doi = "10.1088/1475-7516/2025/09/016",
    journal = "JCAP",
    volume = "09",
    pages = "016",
    year = "2025"
}

@article{Lee:2016vti,
    author = "Lee, Hayden and Baumann, Daniel and Pimentel, Guilherme L.",
    title = "{Non-Gaussianity as a Particle Detector}",
    eprint = "1607.03735",
    archivePrefix = "arXiv",
    primaryClass = "hep-th",
    doi = "10.1007/JHEP12(2016)040",
    journal = "JHEP",
    volume = "12",
    pages = "040",
    year = "2016"
}

@article{Bordin:2018pca,
    author = "Bordin, Lorenzo and Creminelli, Paolo and Khmelnitsky, Andrei and Senatore, Leonardo",
    title = "{Light Particles with Spin in Inflation}",
    eprint = "1806.10587",
    archivePrefix = "arXiv",
    primaryClass = "hep-th",
    doi = "10.1088/1475-7516/2018/10/013",
    journal = "JCAP",
    volume = "10",
    pages = "013",
    year = "2018"
}

@article{Baumann:2020dch,
    author = "Baumann, Daniel and Duaso Pueyo, Carlos and Joyce, Austin and Lee, Hayden and Pimentel, Guilherme L.",
    title = "{The Cosmological Bootstrap: Spinning Correlators from Symmetries and Factorization}",
    eprint = "2005.04234",
    archivePrefix = "arXiv",
    primaryClass = "hep-th",
    doi = "10.21468/SciPostPhys.11.3.071",
    journal = "SciPost Phys.",
    volume = "11",
    pages = "071",
    year = "2021"
}

\end{document}